\documentclass[
showpacs,
 amsmath,amssymb,
prb,
twocolumn,
]{revtex4-2}
\usepackage{graphicx}
\usepackage{dcolumn}
\usepackage{bm}
\usepackage{braket}
\usepackage{color}
\usepackage{accents}
\usepackage{mathrsfs}
\usepackage{bm}
\usepackage{here}
\usepackage{longtable}

\begin{document}
\title{
An odd-frequency Cooper pair around a magnetic impurity}
\author{Shu-Ichiro Suzuki$^{1,2}$}
\author{Takumi Sato$^{3}$}
\author{Yasuhiro Asano$^{3}$}%
\affiliation{
$^{1}$Department of Applied Physics, Nagoya University, Nagoya 464-8603, Japan,\\
$^{2}$ Faculty of Science and Technology, University of Twente, P.O. Box 217,
7500 AE Enschede, The Netherlands,\\
$^{3}$Department of Applied Physics, Hokkaido University, Sapporo 060-8628, Japan.
}
\date{\today}
\begin{abstract}
The Yu-Shiba-Rusinov (YSR) state appears as a bound state of a quasiparticle at a magnetic 
atom embedded in a superconductor.
We discuss why the YSR state has energy below the superconducting gap and why 
the pair potential changes the sign at the magnetic atom.
Although a magnetic atom in a superconductor has been considered as a pair breaker 
since 1960s, we propose an alternative physical picture to explain these reasons.
The analytical expression of the Green's function indicates that a magnetic atom 
converts a spin-singlet $s$-wave Cooper pair into odd-frequency Cooper pairs rather than breaking it
and that the odd-frequency pairing correlations coexist with the YSR states below the gap. 
The relationships among the free-energy density, the amplitudes of pairing correlation functions, 
and the sign change of the pair potential at a magnetic impurity are discussed utilizing the self-consistent 
solution of the Eilenberger equation.
We conclude that the sign change of the pair potential happens only 
when the amplitudes of odd-frequency pairing correlations are dominant at the magnetic impurity.
In the presence of the local $\pi$-phase shift in the pair potential, odd-frequency pairs can 
decrease the free-energy density there because their response to a magnetic field is paramagnetic.
\end{abstract}
\maketitle

\section{Introduction}

The Yu-Shiba-Rusinov (YSR) state~\cite{yu:actphys1965,shiba:ptp1968,rusinov:jetplett1969} 
is a bound state around a magnetic impurity embedded in a superconductor and is 
considered to appear as a result of breaking a spin-singlet Cooper pair by the magnetic moment.
 Superconducting junctions with the YSR states have been 
attracting renewed attention as a possible platform for quantum computer architectures. 
Actually it is known that a chain of magnetic nanoparticles on an $s$-wave superconductor accommodates
Majorana fermions at its ends~\cite{choy:prb2011,klinovaja:prl2013,Nadjerge:science2014}.
Controlling the superconducting subgap state is a necessary element of future quantum 
technology~\cite{menard:natphys2015,scherubl:natphys2020,ebis:prb2015}.

A spin-singlet Cooper pair is formed by two electrons which are time-reversal partner to each other.
Thus a magnetic impurity (a defect breaking time-reversal symmetry) acts as a pair-breaker.
Indeed it is widely accepted that magnetic impurities drastically 
suppress the superconducting transition temperature $T_c$~\cite{abrikosov:jetp1961,maki:book1969}.
Although the formation of YSR state was pointed out in 1960s, there are three unsolved 
issues as listed below.  
(i) This conventional simple picture does not explain why magnetic impurity generates 
the YSR states \textsl{below the gap}. 
(ii) A strong magnetic impurity changes the sign of the pair potential 
around the impurity site~\cite{salkola:prb1997,flatte:prl1997,meng:prb2015,bjornson:prb2017}.
As briefly mentioned in a review paper~\cite{balatsky:rmp2006}, there is no convincing 
explanation for the local $\pi$-phase shift even now.
(iii) It has been unclear the relation between the appearance of the YSR states 
and the suppression of $T_c$.
In this paper, we will show that the existence of odd-frequency Cooper pairs around the 
magnetic impurity provides satisfactory explanation to these issues.

The odd-frequency Cooper pairing is a concept which was introduced by 
Berezinskii~\cite{berezinskii:jetplett1974} to explain the superfluidity in $^3$He.
Although there has been much theoretical work on odd-frequency superconducting states in bulk
systems since 1990s~\cite{kirkpatrik:prl1991,belitz:prb1992,balatsky:prb1992,abrahams:prb1993,coleman:prl1993,abrahams:prb1995,zachar:prl1996,vojta:prb1999},
experimental evidence is still lacking. 
One of authors showed that the spatially uniform odd-frequency 
superconducting order is impossible in single-band metals~\cite{fominov:prb2015}.
However, it turns out that the odd-frequency spin-triplet $s$-wave triplet state can
be realized as an induced subdominant pairing correlation in a rather conventional 
system consisting of a spin-singlet $s$-wave superconductor and a ferromagnet~\cite{bergeret:prl2001}.
Induced odd-frequency pairing correlations have been discussed 
in connection with a subgap quasiparticle appearing at a surface of unconventional superconductors~\cite{buchholz:prb1981,hara:ptp1986,hu:prl1994,tanaka:prl1995,tanaka:jpsj2012}, 
a vortex core \cite{yokoyama:prb2008,tanuma:prl2009}, and an edge of a Majorana nanowire~\cite{asano:prb2013}.
In superconductors having internal degrees of freedom (e.g. sublattices, multi-orbital, and multi-band), 
 a quasiparticle on the Bogoliubov Fermi surface~\cite{agterberg:prl2017} accompanies an
odd-frequency Cooper pair~\cite{kim:jpsj2021}.
Although the odd-frequency pairing correlations around a magnetic impurity were pointed out by 
recent studies~\cite{kuzmanovski:prb2020,perrin:prl2020}, physical phenomena unique to an odd-frequency 
Cooper pair have not been discussed yet.  
The most important property of odd-frequency Cooper pairs 
is that they exhibit a paramagnetic response to an external magnetic 
field~\cite{tanaka:prb2005,asano:prl2011,suzuki:prb2014,asano:prb2015}. 
When the amplitude of an odd-frequency pair is dominant at some place in a superconductor, 
the spatial gradient in the superconducting phase decreases the free-energy there.
Indeed, two of authors showed that the magnetic response of a small unconventional superconductor 
can be paramagnetic at a low temperature~\cite{suzuki:prb2014}. 
We summarize physical consequences of paramagnetic Cooper pairs in
Appendix~\ref{sec:paramagnetic}.

In this paper, we calculate the Green's function around a magnetic impurity in a 
spin-singlet $s$-wave superconductor both analytically and numerically. 
The analytical expression of the anomalous Green's function shows that the magnetic impurity converts
a spin-singlet $s$-wave Cooper pair into an odd-frequency Cooper pair. 
The direct comparison between the normal and the anomalous Green's functions explains well
the coexistence of such an odd-frequency pair and a quasiparticle at the YSR states. 
%
%
Thus the formation of YSR states below the gap is a natural consequence of appearing of an odd-frequency pair. 
As a result of the symmetry conversion of a Cooper pair, 
the amplitude of the spin-singlet $s$-wave pairing correlation decreases 
down to zero and changes the sign with the increase of the amplitude of magnetic moment. 
To explain why the pair potential changes the sign around a magnetic impurity, 
we analyze the free-energy density and the pairing correlation function around the magnetic impurity 
by solving the Eilenberger equation~\cite{eilenberger:zphys1968,larkin:jetp1969} numerically.
The results show that the free-energy density at a magnetic impurity can be  
larger than that in the normal state. 
Such an unusual local state is possible only in a inhomogeneous superconductor. 
 Odd-frequency Cooper pairs increase the free-energy density because
they are thermodynamically unstable under the spatially uniform pair potential.
We also show that the free-energy density at the impurity is decreased by
the $\pi$-phase shift in the pair potential
when odd-frequency Cooper pairs stay more dominantly than even-frequency pairs.
The paramagnetic response of odd-frequency pairs explains naturally the close relationships 
among the appearance of an odd-frequency Cooper pair, 
the local $\pi$-phase shift in the pair potential, and the instability of superconducting state around the impurity.
As an extension of the main conclusions, 
we also discuss the decrease of transition temperature $T_c$ in the presence of a number of magnetic impurities.

This paper is organized as follows. 
In Sec.~II, we summarize the pairing correlations around a magnetic impurity 
embedded in a superconductor in one-dimension.
In Sec.~III, we display numerical results of the pair potential, the local density of states, the pairing 
correlations, and the free-energy density.
We explain why the appearance of odd-frequency pairs decreases $T_c$ in Sec.~IV.
The conclusion is given in Sec.~V.
Throughout this paper, we use the system of units $\hbar=k_B=c=1$, where $k_B$ is the Boltzmann constant 
and $c$ is speed of light.

\section{ Cooper pairs around a magnetic impurity }\label{sec:ana}
Let us consider a spin-singlet $s$-wave superconductor 
in one-dimension where a magnetic impurity is embedded at $x=0$.
The Bogoliubov-de Gennes (BdG) Hamiltonian of a superconductor is given by
\begin{align}
\check{H}_{\mathrm{BdG}}(x)
=& \left[ 
\begin{array}{cc}
\xi_{x}  & \Delta i\hat{\sigma}_2 e^{i\varphi} \\
\Delta (-i)\hat{\sigma}_2 e^{-i\varphi} & - \xi_{x}^\ast
\end{array} \right] +\check{V}(x), \label{eq:hbdg}\\
\xi_{x} = &- \frac{1 }{2m} \frac{d^2}{d x^2} - \epsilon_F,
\end{align}
where $\Delta$ is the uniform pair potential in spin-singlet $s$-wave symmetry, 
$\epsilon_F$ is the Fermi energy, 
$\check{V}$ represents the impurity potential, and $\hat{\sigma}_j$ for $j=1-3$ is the Pauli matrix in spin space. 
The potential of a paramagnetic impurity at $x=0$ is described by 
\begin{align}
\check{V}(x) =& \left[ \begin{array}{cc} \boldsymbol{V} \cdot \hat{\boldsymbol{\sigma}} & 0 \\ 
0 & - \boldsymbol{V} \cdot \hat{\boldsymbol{\sigma}}^\ast \end{array} \right]\, \delta(x).
\label{eq:magimp_def}
\end{align}
The Gor'kov equation reads,
\begin{align}
\left[ i\omega_n - \check{H}_{\mathrm{BdG}}(x) \right]
\check{\mathcal{G}}(x- x^\prime) = \check{1}\, \delta(x-x^\prime),
\end{align}
where $\omega_n=(2n+1) \pi T$ is the Matsubara frequency with $n$ and $T$ being an 
integer number and a temperature, respectively.
The Green's function has a structure
\begin{align}
\check{\mathcal{G}}(x,x^\prime) 
=\left[ \begin{array}{cc}
\hat{g}(x,x^\prime) & \hat{f}(x,x^\prime) \\
-\hat{f}^\ast(x,x^\prime) & -\hat{g}^\ast(x,x^\prime)
\end{array}\right],
\end{align}
because of particle-hole symmetry in the BdG Hamiltonian.
As shown in Appendix~\ref{LSeq}, the Green's function in the presence of an impurity 
can be calculated exactly.
\begin{widetext}
The normal Green's function is shown in Eq.~(\ref{eq:g_m}) in Appendix \ref{LSeq}.
The density of states are calculated from the normal Green's function in the retarded causality
\begin{align}
N(x, \epsilon) = -\frac{1}{4 \pi} \mathrm{Im}\, \mathrm{Tr}\left[ \check{\mathcal{G}}^R_{\epsilon}(x, x) \right]
=-\frac{1}{4 \pi} \mathrm{Im}\, \mathrm{Tr}
\left[ \hat{g}^R_{\epsilon}(x, x) - (\hat{g}^R_{\epsilon})^\ast(x, x) \right] , 
\quad
 \check{\mathcal{G}}^R_{\epsilon}(x, x^\prime) = \left.\check{\mathcal{G}}(x, x^\prime)\right|_{i\omega_n \to \epsilon+i\delta},
\end{align} 
where $\delta$ is a small positive real value. 
The retarded Green's function at $x=x^\prime$ for $0 \leq \epsilon < \Delta$ is calculated as
\begin{align}
\mathrm{Tr}\left[ \check{\mathcal{G}}^R_{\epsilon}(x, x) \right] =& \frac{4 \pi \,N_0\, \epsilon} {\sqrt{  \Delta^2 - \epsilon^2 }} 
\left[ 1 - e^{-2|x|/\xi_0} |\boldsymbol{\gamma}|^2 
\frac{ 2 \Delta^2  + \left\{  \Delta^2(1-|\boldsymbol{\gamma}|^2) + \epsilon^2(1+|\boldsymbol{\gamma}|^2) \right\}
 \cos 2k|x| }{\Delta^2(1-|\boldsymbol{\gamma}|^2)^2 -\epsilon^2(1+|\boldsymbol{\gamma}|^2)^2} \right], 
 \label{eq:gr_xx}
\end{align}
where $N_0$ is the density of states per spin at the Fermi level
and $\boldsymbol{\gamma} =\pi N_0 \, \boldsymbol{V}$.
The first term representing the 
bulk superconducting gap does not contribute to the density of states for $0< |\epsilon| <\Delta$. 
The second term represents a quasiparticle excitation at a magnetic impurity.
It is easy to show that the denominator has poles at $\epsilon = \pm \epsilon_0$ with
\begin{align}
\epsilon_0 = \Delta \frac{1-|\boldsymbol{\gamma}|^2}{1+|\boldsymbol{\gamma}|^2}, \label{eq:e0}
\end{align}
which corresponds to an energy of the YSR state. 
The anomalous Green's function is also calculated as
\begin{align}
\hat{f}(x,x^\prime) =& \frac{\pi N_0 \Delta} {\Omega}
\left[ - \cos k(|x-x^\prime|) e^{-|x-x^\prime|/\xi_0}  
+\frac{e^{-(|x|+|x^\prime|)/\xi_0} |\boldsymbol{\gamma}|^2}{Z}
\left\{2\omega_n^2 (C_++C_-) - \Omega^2(1-|\boldsymbol{\gamma}|^2)C_+\right\} 
\right]  i\hat{\sigma}_2 e^{i\varphi} \nonumber\\
+&\frac{\pi N_0 \Delta\, e^{-(|x|+|x^\prime|)/\xi_0}}{Z}
   2\, i\,  \omega_n  \, |\boldsymbol{\gamma}|^2 \, S_-\, i\hat{\sigma}_2 e^{i\varphi}
-\frac{\pi N_0 \Delta\, e^{-(|x|+|x^\prime|)/\xi_0}}{Z}
\Omega\, (1-|\boldsymbol{\gamma}|^2) S_-  
\boldsymbol{\gamma} \cdot \hat{\boldsymbol{\sigma}} i\hat{\sigma}_2 e^{i\varphi} 
\nonumber\\
+& \frac{\pi N_0 \Delta\, e^{-(|x|+|x^\prime|)/\xi_0}}{Z} 
 i \, \omega_n \,  \left\{ (1+ |\boldsymbol{\gamma}|^2)C_+ + (1- |\boldsymbol{\gamma}|^2)C_- \right\} 
 \boldsymbol{\gamma} \cdot \hat{\boldsymbol{\sigma}} 
\, i \hat{\sigma}_2 e^{i\varphi}, \label{eq:f_m}
\end{align}
with $S_\pm$ and $C_\pm$ in Eq.~(\ref{eq:spmcpm_def}), and $Z$ in Eq.~(\ref{eq:csz_def}).
The similar results were obtained in the previous paper\cite{kuzmanovski:prb2020}.
\end{widetext}

\begin{table*}[t]
\begin{center}
\begin{tabular}{ccccc}
\hline
\null & translational  & local inversion & spin-rotation  & time-reversal \\
\hline
a magnetic impurity & $\times$ & $\times^{(a)}$ & $\times^{(b)}$ & $\times$ \\
\hline
 Born approximation &  $\circ$ & $\circ$ &  $\circ$ &  $\times^{(c)}$ \\
  \hline
 present theory  & $\circ$ & $\circ$ &  $\times^{(d)}$ &  $\times $ \\
  \hline
\end{tabular}
\end{center}
\caption[b]{
Symmetry of the potential for random magnetic impurities in three theoretical models.
The potential of a magnetic impurity in Eq.~(\ref{eq:magimp_def}) breaks translational symmetry, 
inversion symmetry locally, spin-rotation symmetry, and time-reversal symmetry.
Breaking local inversion symmetry enables the parity conversion of the pairing correlation between 
even-parity and odd-parity as indicated by (a). 
Breaking spin-rotation symmetry enables the spin conversion of the pairing correlation between 
spin-singlet and spin-triplet as indicated by (b). 
In the self-consistent Born approximation resulting Eqs.~(\ref{eq:scb_mag}) and (\ref{eq:tc_ag_mag}), 
translational, local inversion, and spin-rotation 
symmetries are restored in the self-energy due to magnetic impurities. 
The renormalization factor changes the sign in the absence of time-reversal symmetry as indicated by (c).
In the present theory resulting Eq.~(\ref{eq:final}), translational and local inversion symmetries 
recovered by averaging of Eq.~(\ref{eq:random2}). 
Since spin-rotation symmetry is broken as indicated by (d), the odd-frequency spin-triplet even-parity 
state can be considered as an intermediate state for scatterings.
}
\label{table1}
\end{table*}
%
The first term stems from the unperturbed Green's function. 
The second term is the pairing correlation at the impurity.
These two terms belong to even-frequency spin-singlet even-parity $s$-wave symmetry class
and are linked to the pair potential 
through the gap equation.
In Table.~\ref{table1}, we summarize symmetries broken by a magnetic impurity.
Eq.~(\ref{eq:magimp_def}) breaks translational symmetry, inversion symmetry in the vicinity of $x=0$, 
spin-rotation symmetry, and time-reversal symmetry. 
As a consequence, a magnetic impurity generates various pairing correlations as shown 
in Eq.~(\ref{eq:f_m}). 
The absence of local inversion symmetry allows the generation of an odd-parity $p$-wave Cooper pair 
from an even-parity $s$-wave pair. Indeed, the third term in Eq.~(\ref{eq:f_m}) is the pairing
correlation belonging to odd-frequency spin-singlet odd-parity symmetry class because 
it is an odd function of $\omega_n$, proportional to $\hat{\sigma}_2$, 
 and antisymmetric under interchanging of $x \leftrightarrow x^\prime$.
The breaking spin-rotation symmetry enables the generation of a spin-triplet Cooper pair 
from a spin-singlet pair.
The last term in Eq.~(\ref{eq:f_m}) represents the pairing correlation belonging to
odd-frequency spin-triplet even-parity symmetry class. 
It is easy to confirm that the last term is symmetric under 
interchanging of $x \leftrightarrow x^\prime$.
As a result of breaking spin-rotation symmetry and local inversion symmetry simultaneously, 
even-frequency spin-tiplet odd-parity pairing correlation appears as indicated by 
the fourth term in Eq.~(\ref{eq:f_m}). 
It is more natural to think that the magnetic impurity converts a spin-singlet $s$-wave Cooper 
pair into a Cooper pair belonging to another symmetry classes rather than destroying it.

Comparing the normal and the anomalous Green's function 
enables us to understand
the close relationship between odd-frequency pairs and quasiparticles in the YSR 
states.
The LDOS derived from the second term of 
Eq.~(\ref{eq:gr_xx}) is calculated as 
\begin{align}
N_{\mathrm{YSR}}&(x, \epsilon)= 
-N_0\, \Delta
\frac{|\boldsymbol{\gamma}|} {1+|\boldsymbol{\gamma}|^2} \nonumber\\
&\times Y(x) \,
\mathrm{Im} \left[ \frac{1}{\epsilon+ i\delta - \epsilon_0} +  \frac{1}{\epsilon+ i\delta + \epsilon_0}\right],
\label{eq:zep_ldos}
\end{align}
for $\epsilon \approx \pm \epsilon_0$, where
 the function
\begin{align}
Y(x)= e^{-2|x|/\xi_0}( \cos^2 kx + |\boldsymbol{\gamma}|^2  \sin^2 kx),
\end{align}
represents $x$ dependence of the Green's function and 
\begin{align}
\frac{1}{\epsilon + i\delta \pm \epsilon_0} = \frac{\mathcal{P}}{\epsilon  \pm \epsilon_0}- i \,\pi \, \delta(\epsilon \pm \epsilon_0),
\end{align}
gives two peaks in the LDOS due to YSR states below the gap.
The last term of Eq.~(\ref{eq:f_m}) indicated by $\hat{f}_{\mathrm{OTE}}$ 
describes the odd-frequency spin-triplet even-parity pairing correlation and is 
calculated to be
\begin{align}
\hat{f}_{\mathrm{OTE}}&(x,x) = - \pi \, N_0\, \Delta
\frac{\boldsymbol{\gamma} \cdot \hat{\boldsymbol{\sigma} }}
 {1+|\boldsymbol{\gamma}|^2} \, i \hat{\sigma}_2 e^{i\varphi} \nonumber\\
 &\times Y(x)
\left[ \frac{1}{i \omega_n - \epsilon_0} +  \frac{1}{i \omega_n+  \epsilon_0}\right]. \label{eq:ote}
\end{align}
The two Green's functions in Eqs.~(\ref{eq:zep_ldos}) and (\ref{eq:ote}) 
have the same $x$ dependence and the same singularity in energy. 
Since the two Green's functions satisfy the Gor'kov equation,
the singularity at $\epsilon= \pm \epsilon_0$ in the normal Green's function in 
Eq.~(\ref{eq:zep_ldos}) and that in the anomalous Green's function in Eq.~(\ref{eq:ote}) 
compensate each other.
The former describes the peaks in the LDOS reflecting the existence of YSR states
and the latter represents the odd-frequency pairing correlation.
Therefore, odd-frequency Cooper pairs and subgap quasiparticles at YSR states coexist with each other. 
As far as we have studied, odd-frequency pairs are almost always associated with subgap quasiparticles.
Thus the appearance of the YSR states below the gap is a direct consequence of the 
generation of the odd-frequency pairing correlations.

The pair potential and the pairing correlation functions are related to each other 
by the gap equation
\begin{align}
\Delta(x) =& T\sum_{\omega_n} \lambda\,   
 \frac{1}{2} \mathrm{Tr}\left[ 
\hat{f}(x,x)(-i \hat{\sigma}_2) \right], 
\end{align}
where $\lambda$ is the strength of the short-range attractive interaction between two electrons.
Tracing in spin space and putting $x^\prime \to x$, the spin-singlet $s$-wave component 
is extracted from Eq.~(\ref{eq:f_m}).
As a result, only the first two terms in Eq.~(\ref{eq:f_m}) contribute to the pair potential.
The gap equation at $x=0$
\begin{align}
\Delta(0) = \pi\,  \lambda\, N_0\,  T\, \sum_{\omega_n}  \frac{\Delta\, \Omega\, (1- |\boldsymbol{\gamma}|^2)}{Z},
\label{eq:scf_delta0}
\end{align}
suggests that
the pair potential at the impurity decreases with the increase of the magnetic moment.
The suppression of $\Delta$ 
can be interpreted as a result of the pair conversion from spin-singlet $s$-wave pair 
to an odd-frequency pair. 
Strictly speaking, an idealistic magnetic impurity characterized by the delta-function 
in Eq.~(\ref{eq:magimp_def}) does not change the sign of the pair potential~\cite{rusinov:jetplett1969}.
But a magnetic impurity with a finite size causes the sign change of the pair potential 
when its magnetic moment is larger than a critical value irrespective of spatial dimension of a superconductor~\cite{salkola:prb1997,flatte:prl1997}.
It has been unclear what causes the sign change of the pair potential~\cite{balatsky:rmp2006}.
As we discussed briefly in the introduction, an odd-frequency pair indicates the paramagnetic response to 
magnetic field and favors the spatial gradient in the superconducting phase.
Therefore, odd-frequency pairs around the magnetic impurity stabilize the local $\pi$-phase shift in the pair 
potential. In the next section, we will check the validity of this conclusion by numerical simulation.

%
\begin{figure*}[tbp]
\begin{center}
  \includegraphics[width=18.0cm]{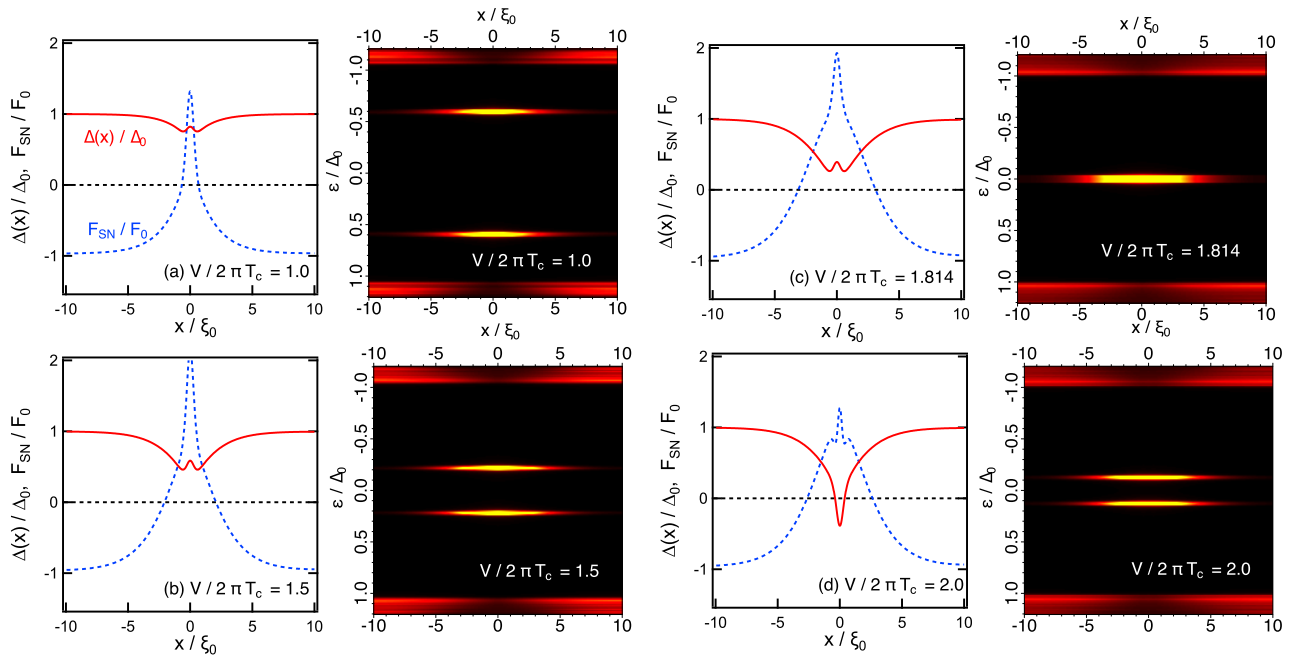}  
\caption{
The numerical results of the pair potential $\Delta(x)$ and the free-energy density $F_{\mathrm{SN}}$, 
and the local density of states $N(x,\epsilon)$  
are plotted for several choices of the amplitude of magnetic moment 
as $|\boldsymbol{V}|/2\pi T_c = 1.0$ in (a), 1.5 in (b), 1.814 in (c), and 2.0 in (d).
The pair potential shown with solid line is normalized to $\Delta_0$ which is the amplitude of the pair potential
in a uniform superconductor at $T=0$.
The free-energy density shown with broken is normalized to $F_0=N_0 \Delta_0^2/2$.
The local density of states is shown on the right panel, where we solve the Eilenberger equation 
for complex energies $\epsilon + i \delta$ with $\delta/2\pi T_c=0.005$.
The size of a magnetic impurity in Eq.~(\ref{eq:x0}) is set as $x_0/ \xi_0=0.5$.}
\label{fig:single}
\end{center}
\end{figure*}
%

\section{ Sign change in pair potential }

In Sec.~\ref{sec:ana}, we assume that the pair potential is uniform in real space.
In this section, we check the validity of our conclusions in the presence of spatial 
variation in the pair potential. Especially, we focus on the effects of odd-frequency pairs 
on the sign change of the pair potential. 
We solve the Eilenberger equation in a superconductor in one-dimension, where 
the potential of a magnetic impurity is described by
\begin{align}
\boldsymbol{V}(x) = \boldsymbol{V} \, e^{-(\frac{x}{x_0})^{2}}, \label{eq:x0}
\end{align}
where $x_0$ represents the spatial range of the impurity potential.
The details of the simulation are shown in Appendix
\ref{eilenberger}.
We numerically calculate the pair potential in Eq.~(\ref{eq:del_scf}) and the local 
density of states (LDOS) in Eq.~(\ref{eq:ldos}). 
In this section, we measure the length and the energy in units of the coherence length $\xi_0= v_F / 2\pi T_c$ and $2\pi T_c$, respectively. 
Here we summarize parameters used in numerical simulation.
The length of the superconductor is fixed at 20$\xi_0$, $x_0$ in Eq.~(\ref{eq:x0}) is 0.5 $\xi_0$, 
and the cut-off energy $\omega_c$ for the Matsubara frequency is 3 $\times 2\pi T_c$.

Figure~\ref{fig:single} shows the pair potential and LDOS around a magnetic impurity 
for several choices of the amplitude of magnetic moment 
$|\boldsymbol{V}|$, where $\Delta_0$ is the amplitude of the pair potential in the bulk at $T=0$. 
 The results at $|\boldsymbol{V}|/2\pi T_c=1.0$ in Fig.~\ref{fig:single}(a) 
 show the slight suppression of the pair potential around a magnetic impurity. 
 The LDOS on the right panel of (a) indicates the 
 existence of the YSR states around the impurity. 
 The energy of the bound state is about $\epsilon = \pm 0.6 \Delta_0$ 
which is relating to the minimum gap size around the impurity. 
For $|\boldsymbol{V}|/2\pi T_c=1.5$ in Fig.~\ref{fig:single}(b), 
a magnetic impurity suppresses the pair potential further. As a result,
the two bound-state energies get closer to the Fermi level as shown in the right panel in (b).
The minimum gap size is $0.45 \Delta_0$ and the energy of a YSR state are
$\pm 0.21 \Delta_0$.
Although the minimum gap size for $|\boldsymbol{V}|/2\pi T_c=1.814$ 
is $0.26 \Delta_0$ in Fig.~\ref{fig:single}(c), the two YSR states exist at almost zero energy. 
When the magnetic moment increases up to $|\boldsymbol{V}|/2\pi T_c=2$, 
the pair potential at the impurity changes the sign and the two bound state energies cross 
as shown in Fig.~\ref{fig:single}(d).
The one-dimensional superconductor with a magnetic impurity is resemble to SFS junction. 
The cross of the Andreev bound state energies at an SFS junction is responsible for 
the transition between the 0-state and the $\pi$-state\cite{rouco:prb2019}.

\begin{figure}[tbp]
\begin{center}
  \includegraphics[width=9cm]{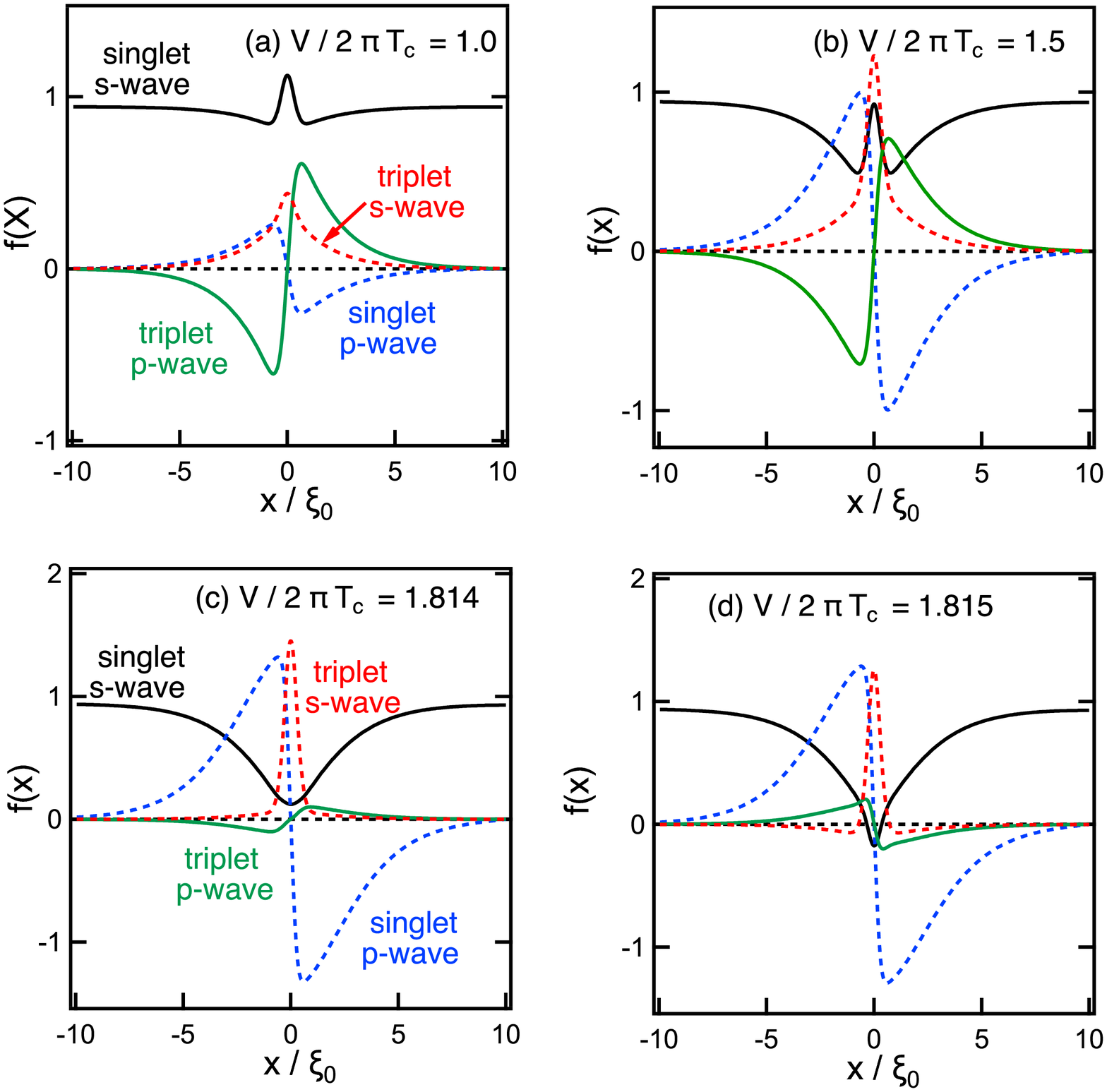}
\caption{
The amplitude of the pairing correlations around the magnetic impurity 
for (a) $|\boldsymbol{V}|/2\pi T_c= 1$,  (b) 1.5, (c) 1.814,  and (d) 1.815.
The solid (broken) lines are the results of even-frequency (odd-frequency) components.
The Matsubara frequency is fixed at $\omega_{n=0}/2\pi T_c =0.1$.
 }
\label{fig:f-single}
\end{center}
\end{figure}

In Fig.~\ref{fig:f-single}, we plot the amplitude of the pairing correlations around the 
magnetic impurity at $\omega_n/2\pi T_c=0.1$, where we choose 
(a) $|\boldsymbol{V}|/2\pi T_c =1.0$, (b) 1.5, (c) 1.814, and (d) 1.815.  
In Fig.~\ref{fig:f-single}(a), spin-singlet $s$-wave component is the most dominant and 
has a small peak at $x=0$, which 
results in a tiny peak in the pair potential at $x=0$ as shown in Fig.~\ref{fig:single}(a).
The amplitude of even-frequency spin-triplet $p$-wave component is the second most dominant.
The odd-frequency spin-triplet $s$-wave and the odd-frequency spin-singlet $p$-wave components
shown with broken lines are subdominant 
everywhere in a superconductor at $|\boldsymbol{V}|/2\pi T_c =1.0$.
The appearance of an odd-frequency 
pair and the spatial variation in the pair potential are related to each other~\cite{asano:prb2014}. 
In this case, the local odd-frequency pairing correlation  
generates the tiny peak at $x=0$ in the pair potential in Fig.~\ref{fig:single}(a).
In addition to this, odd-frequency pairs make the superconducting state unstable locally. 
We calculate the free-energy density in Eq.~(\ref{eq:fsn}) and plot the results 
with a broken line in Fig.~\ref{fig:single}(a). The free-energy density is almost 
equal to $-F_0$ far from the impurity, where $F_0=N_0 \Delta_0^2/2$ is the condensation energy 
of the uniform superconducting state measured from the energy of the normal state.
The negative free-energy means that the superconducting state is more stable than the normal state.
 Although the odd-frequency pairing correlations are subdominant as shown in Fig.~\ref{fig:f-single}(a), 
 the free-energy density becomes positive at $x=0$ in Fig.~\ref{fig:single}(a).
 The positive free-energy at some place does not mean the absence of the pair potential there immediately.
Such a locally destroyed superconductivity cannot be a self-consistent solution of the Eilenberger equation.
To achieve zero pair potential at the impurity, the spin-singlet $s$-wave correlation 
must rapidly become zero in real space. This costs the kinetic energy of the superconducting condensate. 
The nonzero pair potentials under the positive free-energy density are only locally possible in  
inhomogeneous superconductors. 
The total free-energy of such an inhomogeneous superconducting state is lower than that in the normal state.

Under spatially uniform pair potential, 
odd-frequency Cooper pairs are thermodynamically unstable because of their paramagnetic property. 
The results of the free-energy density in Fig.~\ref{fig:single} show that the 
 superconducting state is locally unstable due to the presence of odd-frequency pairs. 
When we increase the magnetic moment to $|\boldsymbol{V}|/2\pi T_c =1.5$, 
the spin-singlet $s$-wave correlation decreases its amplitude slightly as shown in
 Fig.~\ref{fig:f-single}(b). In contrast, the two odd-frequency pairing correlations 
 grow as shown with broken lines.
As a result, the free-energy density around $x=0$ in Fig.~\ref{fig:single}(b) 
increases and becomes larger than that in (a).
In Fig.~\ref{fig:f-single}(c),
we increase the magnetic moment to $|\boldsymbol{V}|/2\pi T_c =1.814$.
The spin-singlet $s$-wave correlation is suppressed drastically at $x=0$ and
the two odd-frequency pairing correlations become dominant locally around the 
impurity as shown with two broken lines. 
In Fig.~\ref{fig:f-single}(d), we increase the magnetic moment only 
slightly to $|\boldsymbol{V}|/2\pi T_c =1.815$.
The spin-singlet $s$-wave component changes the sign around $x=0$, which leads to the 
local sign change of the pair potential. 
The profiles of the two odd-frequency components, on the other hand, 
are insensitive to the slight increase of $|\boldsymbol{V}|$.

\begin{figure}[tbp]
\begin{center}
  \includegraphics[width=9cm]{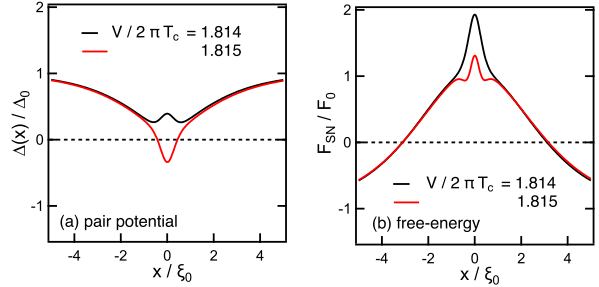}
\caption{
The pair potential (a) and the free-energy density (b) are plotted 
for $|\boldsymbol{V}|/2\pi T_c= 1.814$ and 1.815. 
At $|\boldsymbol{V}|/2\pi T_c= 1.815$,
the pair potential at $x=0$ changes the sign as shown in (a).
As a result of the sign change, the free-energy at $|\boldsymbol{V}|/2\pi T_c= 1.815$ 
can be smaller than that at $|\boldsymbol{V}|/2\pi T_c= 1.814$.
 }
\label{fig:sign_change}
\end{center}
\end{figure}

In Fig.~\ref{fig:sign_change}, we compare the pair potential in (a) and the free-energy density in (b)
at the two values of $|\boldsymbol{V}|/2\pi T_c=1.814$ and 1.815. 
The abrupt sign change of the pair potential in Fig.~\ref{fig:sign_change}(a) 
happens as a result of the finite size effect.
The local sign change in the pair potential causes the drastic decrease of the free-energy density 
 as shown in Fig.~\ref{fig:sign_change}(b). 
The free-energy for $|x|<0.8 \xi_0$ in the presence of the sign change is smaller than that 
in the absence of the sign change. 
As displayed in both Figs.~\ref{fig:f-single}(c) and (d), odd-frequency pairs are
dominant in such places.  
As odd-frequency pairs are paramagnetic, 
the sign change of the pair potential is one of the possible solutions 
to decrease the free-energy density. 
We conclude that odd-frequency pairs generated by a magnetic impurity 
cause the local $\pi$-phase shift in the pair potential near the impurity.

Finally, we show that the overall sign change in spin-triplet $p$-wave component 
Figs.~\ref{fig:f-single}(c) and (d) do not affect the free-energy density.
The pairing correlations are calculated from the parameters $a_\nu$ for $\nu=0-3$ as 
shown in Eqs.~(\ref{eq:def_fcomponent})-(\ref{eq:ldos}). 
Thus the overall sign change of the pairing correlation is derived from 
that of $a_\nu$, (i.e., $a_\nu \to - a_\nu$).
 The induced pairing correlations contribute to 
 the free-energy density through the second term in Eq.~(\ref{eq:fsn}).
 The function $\mathcal{N}_0$ shown in Eq.~(\ref{eq:ldos}) consists of the products 
 of $a_\nu$ and its particle-hole conjugation
$\underline{a}_\nu$. Thus the overall sign change of the spin-triplet $p$-wave
component does not affect the free-energy density because
 $a_\nu\, \underline{a}_\nu$ remains unchanged under $a_\nu$ to $-a_\nu$.

\section{Suppression of $T_c$}
\begin{figure}[tbp]
\begin{center}
  \includegraphics[width=9.0cm]{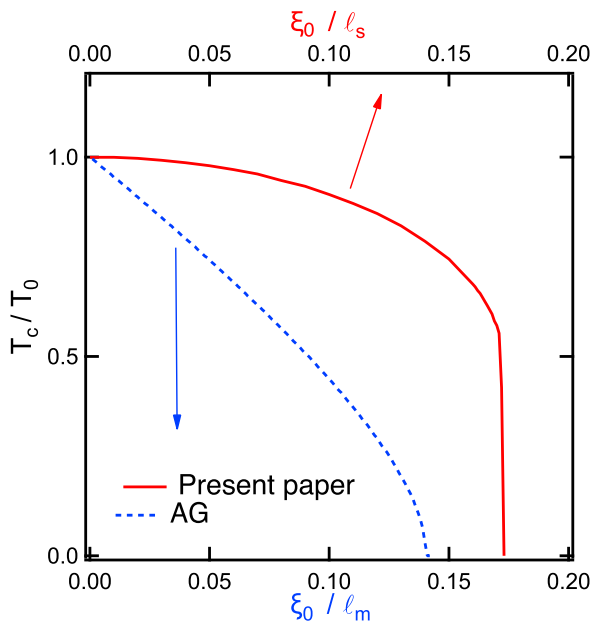}  
\caption{
The theoretical results of transition temperature are plotted as a function 
of the inverse of the mean-free path.
The broken line is the results calculated from the impurity self-energy within the self-consistent Born 
approximation, where we use Eq.~(\ref{eq:scb_mag}).
The solid line is the results of this paper, where 
we use Eq.~(\ref{eq:final}).
It is impossible to compare
$T_c$ in the two results because the horizontal lines are defined in different ways. 
In Eq.~(\ref{eq:scb_mag}), $\ell_{\mathrm{m}} =v_F \tau_{\mathrm{m}}$ is the mean-free path
of an electron on the Fermi level. 
In Eq.~(\ref{eq:final}), on the other hand,
 $\ell_s = \tau_s v_F$ can be interpreted as the mean-free path of a spin-singlet Cooper pair.
}
\label{fig:tc_ana}
\end{center}
\end{figure}

It has been widely accepted that the transition temperature decreases 
with the increase of the magnetic impurity concentration
since 1960s~\cite{abrikosov:jetp1961,maki:book1969}.
We first summarize the outline of these historical papers.
The transition temperature is determined by solving the linearized gap equation
\begin{align}
\Delta = g N_0 2\pi T \sum_{\omega_n>0} F, \label{eq:lgap}
\end{align}
where $F$ is the linearized anomalous Green's function and is
$F=F_{0} = \Delta / |\omega_n|$ in the clean limit.
The anomalous Green's function is renormalized by the self-energy due to impurity scatterings.
For nonmagnetic impurities, the anomalous Green's function is calculated as
\begin{align}
F=& F_{\mathrm{nm}} = \frac{\tilde{\Delta}}{\tilde{\omega}_n}, \\
\tilde{\Delta} =& \Delta \left[1+ \frac{1}{2 \tau_{\mathrm{n}} |\omega_n|}\right], \;
\tilde{\omega}_n = \omega_n \left[1+ \frac{1}{2 \tau_{\mathrm{n}} |\omega_n|}\right],
\end{align}
 where $\tau_{\mathrm{n}}$ is the lifetime of an electron due to scatterings by nonmagnetic impurities.
The relation $F_{\mathrm{nm}}=F_0$ holds true because the renormalization factor 
in the numerator of $F_{\mathrm{nm}}$ cancels that in the denominator. 
As a result, the transition temperature does not change in the presence of nonmagnetic impurities.
For magnetic impurities, however, we find within the self-consistent Born approximation that
\begin{align}
F=& F_{\mathrm{m}} = \frac{\tilde{\Delta}}{\tilde{\omega}_n}= 
\frac{\Delta}{|\omega_n| + 1/\tau_{\mathrm{m}}}, \label{eq:scb_mag}
\end{align}
 where $\tau_{\mathrm{m}}$ is the lifetime of an electron due to scatterings by magnetic impurities.
In Table.~\ref{table1}, we summarize symmetries broken by the self-energy due to magnetic impurities.
Thus translational symmetry, local inversion symmetry, and spin-rotation symmetry 
are restored after averaging. 
The renormalization factor of $\omega_n$ and that of $\Delta$ can be different 
from each other because the magnetic moments of impurities break time-reversal symmetry
\begin{align}
\hat{\sigma}_2\, \boldsymbol{V}\cdot \hat{\boldsymbol{\sigma}}^\ast \, \hat{\sigma}_2 = -  
\boldsymbol{V}\cdot \hat{\boldsymbol{\sigma}}. \label{eq:trs_break}
\end{align} 
The negative sign on the right-hand side is the source of $1/\tau_{\mathrm{m}}$ in Eq.~(\ref{eq:scb_mag}) 
that explains the suppression of $T_c$.
Indeed, the transition temperature is estimated as 
\begin{align}
&\log\left(\frac{T_c}{T_0}\right) \approx 
\psi\left(\frac{1}{2}\right) - \psi\left(\frac{1}{2} + \frac{\xi_0}{\ell_m}\frac{T_0}{T_c}\right),
\label{eq:tc_ag_mag}\\
&\psi\left(\frac{1}{2}\right) - \psi\left(\frac{1}{2} + \frac{x}{2}\right) \nonumber\\
&=
\sum_{n=0}^{\infty} 
\frac{2}{2n +1 + x } -
\sum_{n=0}^{\infty} \frac{2}{2n +1} ,
\end{align}
where $T_0$ is the transition temperature in the clean limit, 
$\ell_{\mathrm{m}}=v_F\, \tau_{\mathrm{m}}$ is the mean free path of an electron due to scatterings 
by magnetic impurities,
and $\psi(x)$ is the digamma function. 
As shown with a broken line in Fig.~\ref{fig:tc_ana}, 
$T_c$ decreases rapidly with the increase of $\xi_0/\ell_{\mathrm{m}}$.
We note that
neither spin-triplet pairing correlations nor odd-frequency pairing correlations 
are considered on the way to Eq.~(\ref{eq:tc_ag_mag}).

Secondly, we discuss how odd-frequency pairs affect the transition temperature.
To do this, we derive the linearized Eilenberger equation which are displayed 
in Eqs.~(\ref{eq:le1}) and (\ref{eq:le2}).
The gap equation in the linearized regime is given in Eq.~(\ref{eq:lg1}).
We consider a pair of the classical trajectories: one is along $\hat{\boldsymbol{k}}$ and 
the other along $-\hat{\boldsymbol{k}}$.
As a result, we obtain following equations which relates the four pairing correlations 
belonging to different symmetry classes,
\begin{align}
& v_F \hat{\boldsymbol{k}} \cdot \nabla  S_-  + 2 \omega_n \, S_+  - 2i \boldsymbol{T}_+ \cdot 
\boldsymbol{V} \nonumber\\
& - 2 \mathrm{sgn}(\omega_n) \,\Delta=0, \label{eq:le_se}\\
&v_F \hat{\boldsymbol{k}} \cdot \nabla \boldsymbol{T}_- 
+ 2 \omega_n \, \boldsymbol{T}_+  - 2i S_+ 
 \boldsymbol{V}=0, \label{eq:le_te}\\
& v_F \hat{\boldsymbol{k}} \cdot \nabla  S_+  + 2 \omega_n \, S_-  - 2i \boldsymbol{T}_- \cdot 
\boldsymbol{V}
=0, \label{eq:le_so}\\
& v_F \hat{\boldsymbol{k}} \cdot \nabla \boldsymbol{T}_+ 
+ 2 \omega_n \, \boldsymbol{T}_-  - 2i S_- 
 \boldsymbol{V}=0,\label{eq:le_to}\\
&S_\pm(\boldsymbol{r}, \hat{\boldsymbol{k}}, \omega_n) = 
a_0(\boldsymbol{r}, \hat{\boldsymbol{k}}, \omega_n) \pm 
a_0(\boldsymbol{r}, -\hat{\boldsymbol{k}}, \omega_n), \\
& \boldsymbol{T}_\pm (\boldsymbol{r}, \hat{\boldsymbol{k}}, \omega_n) = 
 \boldsymbol{a}(\boldsymbol{r}, \hat{\boldsymbol{k}}, \omega_n) \pm 
 \boldsymbol{a}(\boldsymbol{r}, -\hat{\boldsymbol{k}}, \omega_n).
\end{align}
The Riccati's parameter $S_+ (S_-)$ represents the spin-singlet even-parity (odd-parity) pairing 
correlation and $\boldsymbol{T}_+ (\boldsymbol{T}_-) $ 
represents the spin-triplet even-parity (odd-parity) pairing correlation. 
Equation~(\ref{eq:le_se}) describes the relation among 
the three spin-singlet even-parity pairing correlations and the pair potential.
The spatial gradient of the odd-parity correlation $S_-$ generates 
the even-parity component, which 
is a result of breaking local inversion symmetry.
For simplicity, we neglect the effects of breaking local inversion symmetry
and delete the gradient terms in real space in Eqs.~(\ref{eq:le_se})-(\ref{eq:le_to}). 
In this approximation, we implicitly consider the homogeneous correlation functions 
after averaging over the position of magnetic impurities. 
The correlation functions under the approximation recover both translational symmetry and local 
inversion symmetry. 

In the absence of gradient terms in real space,
 $S_-=\boldsymbol{T}_-=0$ is a solution of Eqs.~(\ref{eq:le_so}) and (\ref{eq:le_to}).
The odd-frequency correlation $\boldsymbol{T}_+$ contributes to the spin-singlet 
pairing correlation as a result of breaking spin-rotation symmetry by magnetic impurities.
In what follows, we analyze the two remaining equations in Eqs.~(\ref{eq:le_se}) and (\ref{eq:le_te}) 
in the absence of gradient terms.
%
Eq.~(\ref{eq:le_te}) implies that the magnetic moment 
 generates the odd-frequency spin-triplet pairing correlation $\boldsymbol{T}_+$ 
 from the spin-singlet even-parity correlation $S_+$ in the absence of spin-rotation symmetry.
The Matsubara frequency $\omega_n$ at the second term enables 
the conversion from the odd-frequency 
correlation $\boldsymbol{T}_+$ to the even-frequency correlation $S_+$.
As a result of the conversion, the free-energy density at a magnetic impurity 
increases above zero as shown in Fig.~\ref{fig:single}. 
The free-energy averaged over a whole superconductor increases 
and the transition temperature decreases with the increase of the magnetic impurity concentration.
The superconducting state disappears when the averaged free-energy becomes zero.
 By eliminating the odd-frequency pairing correlation $\boldsymbol{T}_+$ at
 Eqs.~(\ref{eq:le_se}) and (\ref{eq:le_te}),
 we reach an equation only for the spin-singlet even-parity component,
\begin{align}
 \omega_n^2 S_+  +  \boldsymbol{V} \cdot 
\boldsymbol{V} S_+ =
\Delta |\omega_n|. \label{eq:aes}
\end{align}
We assume the properties of random potential 
\begin{align}
\overline{ \boldsymbol{V}(\boldsymbol{r}) }=0, \quad 
\overline{ \boldsymbol{V}(\boldsymbol{r}) \boldsymbol{V}(\boldsymbol{r}) }=
\frac{1}{\tau_s^2}
\label{eq:random2}
\end{align}
under ensemble averaging, where $1/\tau_s$ can be interpreted as the \textsl{lifetime 
of a spin-singlet Cooper pair} in the presence of magnetic impurities.
This can be confirmed by changing $i\omega_n \to - \partial_\tau \to -i \partial_t$ 
 in Eq.~(\ref{eq:aes}). We find that $\tau_s$ gives the characteristic time scale
 of the equation.   
We finally reach a solution of 
\begin{align}
S_+ =& \frac{\Delta \, |\omega_n|}{\omega_n^2 + 1/\tau_s^2}.\label{eq:final}
\end{align} 
The scatterings by magnetic impurities remove 
the singularity at $\omega_n=0$ at the denominator.
This explains the decrease of $T_c$ when we 
substitute $F=S_+$ into Eq.~(\ref{eq:lgap}). 
In Fig.~\ref{fig:tc_ana}, we show $T_c$ calculated by using Eq.~(\ref{eq:final})
with a solid line, where the horizontal axis is $\xi_0/ \ell_s$ with 
$\ell_s=\tau_s v_F$ meaning the mean-free path of a 
spin-singlet $s$-wave Cooper pair under magnetic impurities.
The calculated results show that $T_c$ decreases with the increase of scatterings by magnetic impurities.
This phenomenon can be understood from two different viewpoints because 
magnetic impurities simultaneously generate odd-frequency Cooper pairs and subgap quasiparticles.
The viewpoint from a Cooper pair shows that the odd-frequency pair increases the free-energy locally.
The viewpoint from a quasiparticle suggests that the occupation of the YSR states
below the Fermi level reduces the condensation energy.
We conclude that these two facts make the superconducting state unstable and decrease $T_c$.

%
%

\section{Conclusion}

We have studied the properties of superconducting state around a magnetic impurity 
embedded in a conventional superconductor.
The analytical results of the anomalous Green's function show that 
 the magnetic impurity converts a spin-singlet $s$-wave Cooper pair into odd-frequency Cooper pairs. 
Comparing the normal and the anomalous Green’s function explains 
the coexistence of odd-frequency Cooper pairs and quasiparticles at the 
Yu-Shiba-Rusinov (YSR) states.
We conclude that the formation of the YSR states below the gap 
is a direct consequence of the appearance of an odd-frequency Cooper pair. 

The numerical results of the free-energy density show that 
the superconducting states around a magnetic impurity are 
thermodynamically unstable because of the paramagnetic property of odd-frequency Cooper pairs. 
This fact explains naturally the remaining open issues listed in the introduction.
The sign of the pair potential changes around an impurity with sufficiently large magnetic moment 
because odd-frequency pairs are dominant around such a strong magnetic impurity.
On the basis of obtained results, we also proposed an alternative scenario to explain
the suppression of the transition temperature in the presence of many magnetic impurities.

\begin{acknowledgments}
The authors are grateful to A.~A.~Golubov, Y.~Tanaka, Ya.~V.~Fominov and S.~Hoshino 
for useful discussion.
This work was supported by JSPS KAKENHI (No. JP20H01857) and 
JSPS Core-to-Core Program (No. JPJSCCA20170002). 
S.-I.~S. is supported
by Grant-in-Aid for JSPS Fellows (JSPS KAKENHI
grant Number JP19J02005) and by Overseas Research Fellowships by JSPS. 
S.-I.~S. acknowledges the hospitality at the University of Twente.
T.~S. is supported in part by the establishment of university 
fellowships towards the creation of science technology innovation from the 
Ministry of Education, Culture, Sports, Science, and Technology (MEXT) of Japan.
\end{acknowledgments}

\appendix
\begin{widetext}

\section{Paramagnetic Cooper pairs}\label{sec:paramagnetic}
The superconducting condensate can be described phenomenologically by 
the macroscopic wave function
\begin{align}
\psi(\boldsymbol{r}) = \sqrt{n_{\mathrm{S}}(\boldsymbol{r})}\, e^{i \theta(\boldsymbol{r})},
\end{align}
where $n_{\mathrm{S}}$ is the density of Cooper pairs and $\theta$ is the phase of the condensate.
The flux quantization is derived from the singlevaluedness of this wave function.
The Josephson effect is explained as the tunnel effect between two superconductors 
characterized by such wave functions. 
The energy of the condensate can be calculated in terms of the macroscopic wave function  
\begin{align}
E=& \int d\boldsymbol{r}\, \frac{\hbar^2}{2m}
\left\{ \left(\nabla + i \frac{e}{\hbar c} \boldsymbol{A} \right)
\psi^\dagger(\boldsymbol{r}) \right\} \left\{ \left(\nabla - i \frac{e}{\hbar c} \boldsymbol{A} \right) \psi(\boldsymbol{r}) \right\},\\
=& \int d\boldsymbol{r}\, \frac{\hbar^2 }{2m}
\left\{ \frac{(\nabla n_{\mathrm{S}})^2}{4n_{\mathrm{S}}} + n_{\mathrm{S}} \left(\nabla \theta - \frac{e}{\hbar c} \boldsymbol{A} \right)^2 \right\}.
\label{eq:e_density}
\end{align}
The first term in Eq. ~(\ref{eq:e_density}) represents the kinetic energy of the condensate and
the second term means the elastic energy of the superconducting phase.
Since $n_{\mathrm{S}}>0$, both the spatial gradient of the density and the spatial gradient of the phase 
increase the energy of the condensate, which describes the rigidity of the superconducting state. 
Therefore, both the pair density and the phase 
are uniform at the ground state in the absence of a magnetic field.
The electric current can be described by
\begin{align}
\boldsymbol{j} = &\frac{e \,\hbar}{2im} \left[ 
\psi^\dagger(\boldsymbol{r}) \left(\nabla - i \frac{e}{\hbar c} \boldsymbol{A} \right) \psi(\boldsymbol{r})
- \left(\nabla + i \frac{e}{\hbar c} \boldsymbol{A} \right)\psi^\dagger(\boldsymbol{r}) \; 
\psi(\boldsymbol{r})
\right]
= \frac{e \, \hbar \,n_{\mathrm{S}}}{m} \nabla \theta - \frac{n_{\mathrm{S}}\, e^2  }{m c} \boldsymbol{A}.  \label{eq:ja}
\end{align} 
Together with the Maxwell equation
$\nabla \times \boldsymbol{H} = \frac{4 \pi}{c} \boldsymbol{j}$, 
we obtain 
the equation for a magnetic field in a superconductor
\begin{align}
\nabla^2 \boldsymbol{H} - \frac{4\pi \, n_{\mathrm{S}} e^2}{m \, c^2} \, \boldsymbol{H} =0. \label{eq:london}
\end{align}
The London's length $\lambda_{\mathrm{L}} = \sqrt{m\, c^2/ 4\pi \, n_{\mathrm{S}}\, e^2}$  
characterizes the spatial variation of a magnetic field.
Equation (\ref{eq:london}) represents the Meissner screening effect of a magnetic field.
The dumping of a magnetic field into a superconductor is described by the negative sign 
at the second term on the last line in Eq.~(\ref{eq:ja}).
The argument above is valid when the pair density $n_{\mathrm{S}}$ is positive everywhere 
in a superconductor.

 Let us assume that the pair density is negative locally at a finite area around $\boldsymbol{r}=\boldsymbol{r}_0$ and 
 discuss the physical consequence of $n_{\mathrm{S}}(\boldsymbol{r}_0)<0$.
The second term in Eq.~(\ref{eq:e_density}) suggests that 
large gradient of the phase and the penetration of magnetic field is necessary 
to decrease the energy of the condensate. 
Namely the condensate with the "negative pair density" is paramagnetic.  
Equation (\ref{eq:london}) with negative $n_{\mathrm{S}}$ suggests also that a magnetic field 
can penetrate into such a paramagnetic superconductor~\cite{tanaka:prb2005,asano:prl2011}.
Such local area around $\boldsymbol{r}_0$ may be no longer superconductive 
because the phase $\theta$ fluctuates easily from a constant value. 
Thus the pair density $n_{\mathrm{S}}$ must be positive to realize the stable homogeneous superconducting ground 
state both electromagnetically and thermodynamically.
However, "pair density" can be negative locally in the presence of an odd-frequency pair. 

A surface of a superconductor breaks inversion symmetry locally. 
As a result of breaking inversion symmetry, 
odd-parity (even-parity) pairing correlations appear at a surface of even-parity (odd-parity) 
superconductor.
Such induced pairing correlations belong to odd-frequency symmetry class
because the surface does not change spin of a Cooper pair.
In what follows, we demonstrate that the pair density can be negative at such a surface
by using an analytical solution of the Eilenberger equation, 
\begin{align}
&\hbar v_F {\hat{\boldsymbol{k}}} \cdot \nabla \hat{g} + [H, \hat{g}]=0, \\
& H= \left[ \begin{array}{cc} \omega_n & \Delta(\boldsymbol{r}, \hat{\boldsymbol{k}}) \\
-s_s \undertilde{\Delta}(\boldsymbol{r}, \hat{\boldsymbol{k}}) & -\omega_n \end{array}\right], \quad
\hat{g}(\boldsymbol{r}, \hat{\boldsymbol{k}}, \omega_n) = \left[ \begin{array}{cc} 
g(\boldsymbol{r}, \hat{\boldsymbol{k}}, \omega_n)  & f(\boldsymbol{r}, \hat{\boldsymbol{k}}, \omega_n)  \\
-s_s \undertilde{f}(\boldsymbol{r}, \hat{\boldsymbol{k}}, \omega_n)  & -g(\boldsymbol{r}, \hat{\boldsymbol{k}}, \omega_n) \end{array}\right].
\end{align}
Here we have reduced to $2\times 2$ particle-hole space by extracting one spin sector of the 
Bogoliubov-de Gennes (BdG) Hamiltonian.
The pair potential obeys the symmetry relation
\begin{align}
\Delta(\boldsymbol{r}, -\hat{\boldsymbol{k}})= \left\{ 
\begin{array}{rl} \Delta(\boldsymbol{r}, \hat{\boldsymbol{k}}) & \text{singlet} \; s_s=-1 \\
-\Delta(\boldsymbol{r}, \hat{\boldsymbol{k}}) & \text{triplet} \; s_s=1, \end{array}\right.
\end{align} 
which is drived from the Fermi-Dirac statistics of electrons.
The Eilenberger equation can be decomposed into three equations~\cite{asano:prb2014},
\begin{align}
v_F \hat{\boldsymbol{k}} \cdot \nabla g =& 2\Delta \, f_S, \quad 
v_F \hat{\boldsymbol{k}} \cdot \nabla f_B = -2 \omega_n\, f_S, \quad
v_F \hat{\boldsymbol{k}} \cdot \nabla f_S = 2( \Delta \, g - \omega_n\, f_B), \label{eq:eilen2}
\end{align}
with
\begin{align}
f_B= & \frac{1}{2}(f - s_s \undertilde{f}), \quad  f_S=  \frac{1}{2}(f + s_s \undertilde{f}).
\end{align}
The quasiclassical Green's functions satisfy the normalization condition
\begin{align}
g^2-s_s \undertilde{f}\, f = g^2 + f_B^2 - f_S^2=1.
\end{align}
In a homogeneous superconductor, we obtain the solution
as
\begin{align}
g= \frac{\omega_n}{\Omega}, \quad f_B= \frac{\Delta(\hat{\boldsymbol{k}})}{\Omega}, \quad f_S=0, 
\quad \Omega=\sqrt{\omega_n^2+\Delta^2(\hat{\boldsymbol{k}})}. \label{eq:gf_bulk}
\end{align}
Thus $f_B$ is interpreted as bulk component of pairing correlation which contributes to the pair potential
through the gap equation 
\begin{align}
\Delta(\boldsymbol{r}, \hat{\boldsymbol{k}}) = T\sum_{\omega_n}\, \int \frac{d \hat{\boldsymbol{k}}^\prime}{S_d} 
\lambda(\hat{\boldsymbol{k}}, \hat{\boldsymbol{k}}^\prime) f(\boldsymbol{r}, \hat{\boldsymbol{k}}^\prime, \omega_n),
\label{eq:gap}
\end{align}
where $\lambda$ represents an attractive interaction between two electrons.
The second equation in Eq.~(\ref{eq:eilen2}) represents the symmetry relationship between $f_B$ and $f_S$. 
Since $\hat{\boldsymbol{k}}$ is an odd-parity function 
and $\omega_n$ is an odd in Matsubara frequency, $f_S$ belongs to the opposite parity and opposite frequency 
symmetry class to $f_B$.
Therefore $f_S$ is considered as an induced pairing component due to the spatial gradient of $f_B$.
Although the anomalous Green's function $f$ in Eq.~(\ref{eq:gap}) consists of both $f_B$ and $f_S$, 
only the bulk component $f_B$ contributes to the pair potential. 
Since $f_S$ is an odd-function of $\omega_n$, the summation of $f_S$ over $\omega_n$ vanishes. 
The Meissner kernel for the quasiclassical Green's function is described as~\cite{higashitani:prb2014}
\begin{align}
{j}_\mu =& - \frac{ 2 n e^2}{m c} \mathcal{Q}_{\mu, \nu} {A}_\nu, \\
\mathcal{Q}_{\mu, \nu} = & d \pi T \sum_{\omega_n}  \int \frac{d \hat{\boldsymbol{k}}}{S_d} 
\hat{k}_\mu\, \hat{k}_\nu\, \partial_{\omega_n} g(\boldsymbol{r}, \hat{\boldsymbol{k}}, \omega_n)=  d \pi T \sum_{\omega_n>0}  \int \frac{d \hat{\boldsymbol{k}}}{S_d} \hat{k}_\mu\, \hat{k}_\nu\,
(f_B^2-f_S^2) \, \partial_{\omega_n} \log \frac{1+ g}{1-g}. \label{eq:q_el}
\end{align}
The last expression suggests that the odd-frequency pairing correlation decreases $\mathcal{Q}_{\mu, \nu}$.
By putting the results in Eq.~(\ref{eq:gf_bulk}) for a spin-singlet $s$-wave superconductor 
into $\mathcal{Q}$, it is possible to 
recover the results of~\cite{agd}
\begin{align}
\mathcal{Q}_{\mu, \nu} =  \pi\,  T\sum_{\omega_n} \frac{\Delta^2}{(\omega_n^2 + \Delta^2)^{3/2}} \delta_{\mu, \nu}. 
\end{align}
The product of $n\, \mathcal{Q}_{\mu, \mu}$ is often referred to as pair density.

As an example of inhomogeneous superconducting states, we consider the condensate near the surface of 
a two-dimensional $p$-wave superconductor. The pair potential is described as
\begin{align}
\Delta(x) = \Delta \cos \theta \tanh \left(\frac{x}{\xi_0}\right), \quad \hat{k}_x= \cos\theta, \quad \hat{k}_y=\sin\theta, \quad \xi_0=\frac{v_F}{\Delta \cos\theta},
\end{align}
where we assume that a surface is at $x=0$ and a $p$-wave superconductor occupies $x>0$, 
the superconducting state is uniform in the $y$ direction, and 
$\xi_0$ is the coherence length.
The solution of Eq.~(\ref{eq:eilen2}) can be obtained as~\cite{schopohl:arxiv1998}
\begin{align}
g(x, \theta, \omega_n) =& \frac{\omega_n}{\Omega_\theta} + \frac{ \Delta^2 \cos^2\theta}{2\omega_n \Omega_\theta}
\cosh^{-2} \left(\frac{x}{\xi_0}\right), \quad
f_B(x, \theta, \omega_n)= \frac{ \Delta \cos \theta}{ \Omega_\theta} \tanh \left(\frac{x}{\xi_0}\right), \\
f_S(x, \theta, \omega_n) =& - \frac{ \Delta^2 \cos^2\theta}{2\omega_n \Omega_\theta}
\cosh^{-2} \left(\frac{x}{\xi_0}\right), \quad \Omega_\theta = \sqrt{\omega_n^2 + \Delta^2 \cos^2\theta}.
\end{align}
 The bulk component $f_B(x, \theta+\pi, \omega_n)=-f_B(x, \theta, \omega_n)$ is odd-parity, 
whereas the surface component $f_S$ is even-parity because of $\cos^2(\theta+\pi) = \cos^2\theta$.
The gap equation in Eq.~(\ref{eq:gap}) is described as
\begin{align}
\Delta(x) = &  T\sum_{\omega_n}\, \int_0^{2\pi} \frac{d\theta^\prime}{2\pi} (2\lambda\, \cos\theta \, \cos\theta^\prime) 
\left[f_B(x,\theta^\prime,\omega_n)+ f_S(x,\theta^\prime,\omega_n)\right],\\
= &\Delta\, \cos\theta\, \tanh \left(\frac{x}{\xi_0}\right) \lambda T\sum_{\omega_n} 
\frac{1}{\sqrt{\omega_n^2+\Delta^2}}.
\end{align}
On the way to the second line, we approximately neglect $\theta$ dependence of $\Omega_\theta$ 
and that of $\xi_0$ 
in $f_{\mathrm{B}}$.
The amplitude of $f_S$ increases with the decrease of $\omega_n$ at its denominator and 
can be larger than the amplitude of $f_B$ for $\omega_n \ll \Delta$. 
The resulting response kernel 
\begin{align}
\mathcal{Q}_{\mu, \nu} \approx \delta_{\mu, \nu}\, \pi T\sum_{\omega_n}\, \int_0^{2\pi} \frac{d\theta}{2\pi}
\frac{\Delta^2 \cos^2\theta}{(\omega_n^2+\Delta^2\cos^2\theta)^{3/2}} 
\left[ 1- 
\frac{\Delta^2\cos^2\theta}{2\omega_n^2} \cosh^{-2} \left(\frac{x}{\xi_0}\right)
\right],
\end{align}
 can be negative for a low temperature $T  \ll T_c$ near 
the surface $0<x < \xi_0$.
The paramagnetic response at a surface of unconventional superconductor was pointed out for 
a $d$-wave superconductor~\cite{higashitani:jpsj1997,walter:prl1998}. 
To make clear the details of paramagnetic effect theoretically, 
 analysis beyond the linear response is necessary.
In Refs.~\cite{suzuki:prb2014,suzuki:prb2015}, the pair potential and a magnetic filed 
are determined self-consistently with each other in a small unconventional superconductor.
A small $p$-wave superconducting disk shows paramagnetic response to a magnetic filed 
at a low temperature. 
Even-frequency pairs stabilize $p$-wave superconductivity in the bulk
and odd-frequency pairs exhibit the paramagnetic response at the surface.

 Finally, we briefly explain the coexistence of an odd-frequency pair and a quasiparticle 
 at the Andreev bound state at a surface. After applying $i\omega_n \to \epsilon +i \delta$, 
 the local density of states for $|\epsilon| < \Delta $ calculated from 
 the second term of the normal Green's function becomes 
\begin{align}
\frac{N(x, \epsilon)}{N_0} = \mathrm{Re} \int \frac{d\theta}{2\pi} 
i \frac{ \Delta^2 \cos^2\theta}{2 (\epsilon + i \delta) \sqrt{\Delta^2\cos^2\theta - \epsilon^2} }
\cosh^{-2} \left(\frac{x}{\xi_0}\right) \to \delta(\epsilon)\, \frac{\Delta}{\pi} \cosh^{-2} \left(\frac{x}{\xi_0}\right).
\end{align} 
The peak of the local density of states at zero energy reflects the 
existence of a quasiparticle at the surface Andreev bound state.

\section{Lippmann-Schwinger equation}\label{LSeq}

The Lippmann-Schwinger equation relates
the Green's function in the presence of perturbations $\check{V}$
to the Green's function in the absence of perturbation as
\begin{align}
\check{\mathcal{G}}(\boldsymbol{r}, \boldsymbol{r}^\prime) 
= \check{\mathcal{G}}^{(0)}(\boldsymbol{r}, \boldsymbol{r}^\prime) 
+ \int d \boldsymbol{r}_1\; \check{\mathcal{G}}^{(0)}(\boldsymbol{r}, \boldsymbol{r}_1) \,
\check{V}(\boldsymbol{r}_1) \,  \check{\mathcal{G}}^(\boldsymbol{r}_1, \boldsymbol{r}^\prime).
\end{align}  
When we consider an impurity potential $\check{V}(\boldsymbol{r}) = \check{V} \delta(\boldsymbol{r})$, the 
equation becomes
\begin{align}
\check{\mathcal{G}}(\boldsymbol{r}, \boldsymbol{r}^\prime) 
= \check{\mathcal{G}}^{(0)}(\boldsymbol{r}, \boldsymbol{r}^\prime) 
+ \check{\mathcal{G}}^{(0)}(\boldsymbol{r}, 0)\, \check{V} \,\check{\mathcal{G}}(0, \boldsymbol{r}^\prime). \label{gi1}
\end{align}  
By putting $\boldsymbol{r}=0$ into the equation, the equation 
\begin{align}
\check{\mathcal{G}}(0,\boldsymbol{r}^\prime) 
= \check{\mathcal{G}}^{(0)}(0,\boldsymbol{r}^\prime) 
+ \check{\mathcal{G}}^{(0)}(0,0)\, \check{V}  \,\check{\mathcal{G}}(0,\boldsymbol{r}^\prime)
\end{align}  
has a closed form. Substituting the solution
\begin{align}
\check{\mathcal{G}}(0, \boldsymbol{r}^\prime) 
= \left[ 1 - \check{\mathcal{G}}^{(0)}(0,0) \check{V} \right]^{-1} \check{\mathcal{G}}^{(0)}(0, \boldsymbol{r}^\prime)
\end{align}  
 into Eq.~(\ref{gi1}), we obtain
\begin{align}
\check{\mathcal{G}}(\boldsymbol{r}, \boldsymbol{r}^\prime) 
= \check{\mathcal{G}}^{(0)}(\boldsymbol{r}, \boldsymbol{r}^\prime) 
+ \check{\mathcal{G}}^{(0)}(\boldsymbol{r}, 0) \, \check{V} \,  
\left[ 1 - \check{\mathcal{G}}^{(0)}(0,0) \check{V} \right]^{-1} \check{\mathcal{G}}^{(0)}(0,\boldsymbol{r}^\prime). \label{gi2}
\end{align}

In this paper, we consider a spin-singlet $s$-wave superconductor in one-dimension 
as described by Eq.~(\ref{eq:hbdg}).
The unperturbed Green's function in the Matsubara representation is given by 
\begin{align}
\check{\mathcal{G}}^{(0)}(x,x^\prime) 
%
=&
-\frac{\pi N_0}{\Omega} \check{U}^\dagger
\left[ 
\begin{array}{cc}
i\omega_n C_0 -\Omega S_0 & \Delta C_0  \\ \Delta C_0
& i\omega_nC_0 +\Omega S_0 \end{array}\right] e^{-|x-x^\prime|/\xi_0}\check{U}, 
\label{g0}
\end{align}
with 
\begin{align}
\Omega=& \sqrt{\omega_n^2 + \Delta^2}, \quad
k_\pm = k \left(1 \pm i \frac{\Omega}{2\epsilon_F}\right), \quad 
\check{U} = \left[\begin{array}{cc} 1 &  0 \\ 0 & i\hat{\sigma}_2 e^{i\varphi}\end{array}\right],\\
C_0 =& \cos(k|x-x^\prime|), \quad S_0 = \sin(k|x-x^\prime|), \quad 
\xi_0= \frac{2\epsilon_F}{\Omega k_F},
\end{align}
where $\hat{\sigma}_j$ for $j=1-3$ is the Pauli's matrix in spin space and $k$ denotes the Fermi wavenumber.

The potential of a nonmagnetic impurity is described by 
\begin{align}
\check{V} = \left[ \begin{array}{cc} V_0 \hat{\sigma}_0 & 0 \\ 0 & - V_0 \hat{\sigma}_0 \end{array} \right]
= V_0 \hat{\tau}_3, \label{eq:nonmag}
\end{align}
where $\hat{\tau}_j$ for $j=1-3$ is the Pauli's matrix in particle-hole space.
The Green's function in the presence of a nonmagnetic impurity is calculates as
\begin{align}
\check{\mathcal{G}}&(x,x^\prime) 
=
\check{\mathcal{G}}^{(0)}(x,x^\prime) 
-\frac{\pi N_0}{\Omega}e^{-(|x|+|x^\prime|)/\xi_0} \frac{\gamma_0}{(1+\gamma_0^2)} 
\nonumber \\
&\times  \check{U}^\dagger 
\left[ 
\begin{array}{cc}
i\omega (S_+ - \gamma_0 C_+) + \Omega(C_+ +\gamma_0 S_+)  &
\Delta (S_+ - \gamma_0 C_+) \\ \Delta (S_+ - \gamma_0 C_+) &
i\omega (S_+ - \gamma_0 C_+) - \Omega(C_+ +\gamma_0 S_+)  
\end{array}\right]\check{U},\\
&C_\pm = \cos\left\{k(|x|\pm|x^\prime|)\right\}, \quad S_\pm = \sin\left\{k(|x|\pm|x^\prime|)\right\}, 
\quad \gamma_0=\pi N_0 V_0. \label{eq:spmcpm_def}
\end{align}
The anomalous Green's function results in
\begin{align}
\hat{f}(x, x^\prime) = - \frac{ \pi N_0 \Delta}{\Omega} \, i \hat{\sigma}_2\, 
\left[ C_0 \, e^{-|x-x^\prime|/\xi_0} + \frac{\gamma_0}{1+\gamma_0^2} \left( S_+ - \gamma_0\, C_+ \right) 
e^{-(|x|+|x^\prime|)/\xi_0} \right].
\end{align}
The anomalous Green's function $\hat{f}(x,x^\prime)$ consists only of a spin-singlet even-parity Cooper pair
because it remains unchanged under $x \leftrightarrow x^\prime$.
The normal Green's function becomes
\begin{align}
\mathrm{Tr}\left[ \check{\mathcal{G}}^R_{\epsilon}(x, x) \right] = \frac{-4 \pi i \,N_0\, \epsilon} {\sqrt{ \epsilon^2 - \Delta^2}} 
\left[ 1+ \frac{\gamma_0}{1+\gamma_0^2} \left( \sin 2 k |x|  - \gamma_0\, \cos 2 k|x| \right) 
e^{-2|x|/\xi_0}\right],
\end{align}
which is always real value for $\epsilon < \Delta$.

In the presence of a magnetic impurity,  
the normal Green's function is calculated to be
\begin{align}
\hat{g}(x,x^\prime) =&  \frac{\pi N_0}{\Omega}
\left[ 
- e^{-|x-x^\prime|/\xi_0} (i\omega_n C_0 - \Omega S_0) \right. \nonumber\\
&-\frac{e^{-(|x|+|x^\prime|)/\xi_0}}{Z}|\boldsymbol{\gamma}|^2
\left[ 2i\omega_n \Delta^2 C_- 
+ 
\left\{ \Delta^2(1-|\boldsymbol{\gamma}|^2) - \omega_n^2(1+|\boldsymbol{\gamma}|^2) \right\}
(i\omega_n C_+ - \Omega S_+) \right]  \nonumber\\
&+ \frac{e^{-(|x|+|x^\prime|)/\xi_0}}{Z} \boldsymbol{\gamma} \cdot \hat{\boldsymbol{\sigma}}
\Omega 
\left.\left[ \Delta^2(1-|\boldsymbol{\gamma}|^2) C_- 
+  i \omega_n (i\omega_n C_+ - \Omega S_+)(1+|\boldsymbol{\gamma}|^2)
\right] \right],\label{eq:g_m}\\
Z=& \Delta^2(1-|\boldsymbol{\gamma}|^2)^2 + \omega_n^2(1+|\boldsymbol{\gamma}|^2)^2, \quad \boldsymbol{\gamma}=\pi N_0 \boldsymbol{V}. \label{eq:csz_def}
\end{align}
where $S_\pm$ and $C_\pm$ are given in Eq.~(\ref{eq:spmcpm_def})
The anomalous Green's function is displayed in Eq.~(\ref{eq:f_m}).

\section{Eilenberger equation}\label{eilenberger}

To study the pairing correlations around a magnetic impurity, we solve the Eilenberger equation 
numerically.
The Matsubara Green function can be decomposed into the Riccati parameters as,
\begin{align}
\check{\mathcal{G}}(\boldsymbol{r}, \hat{\boldsymbol{k}}, i\omega_n)
=&\left(\begin{array}{cc}
 \hat{\mathcal{G}} &
 \hat{\mathcal{F}} \\
 \underline{\hat{\mathcal{F}}} &
 -\underline{\hat{\mathcal{G}}}
\end{array}\right)
 = \left(\begin{array}{cc}
 {\hat{\mathcal{N}}} &
 {\hat{0}} \\
 {\hat{0}} &
 \underline{\hat{\mathcal{N}}}
\end{array}\right)
\left(\begin{array}{cc}
 \mathrm{sgn}(\omega_n)\, (1-\hat{a}\underline{\hat{a}}) & 2 \hat{a} \\
 2\underline{\hat{a}} &
-\mathrm{sgn}(\omega_n)\, ( 1-\underline{\hat{a}}\hat{a})
\end{array}\right),\label{ricatti_m}\\
&\hat{a}=\hat{a}(\boldsymbol{r}, \hat{\boldsymbol{k}}, i\omega_n), \qquad
\underline{\hat{a}}=\underline{\hat{a}}(\boldsymbol{r},\hat{\boldsymbol{k}}, i\omega_n),\label{a_def_matsubara}\\
&\hat{\mathcal{N}} = (1+ \hat{a} \underline{\hat{a}})^{-1},
\qquad \underline{\hat{\mathcal{N}}} = (1+ \underline{\hat{a}}\hat{a} )^{-1}.\label{n_def}
\end{align}
The Riccati parameters can be decomposed into four components as
\begin{align}
\hat{a}(\boldsymbol{r}, \hat{\boldsymbol{k}}, i\omega_n)
=& a_0(\boldsymbol{r}, \hat{\boldsymbol{k}}, i\omega_n) +
\boldsymbol{a}
(\boldsymbol{r}, \hat{\boldsymbol{k}}, i\omega_n) \cdot \hat{\boldsymbol{\sigma}},
\end{align}
where $a_0$ is the spin-singlet component and $a_j$ for $j=1-3$ are three spin-triplet components. 
They obey the symmetry relations 
\begin{align}
a_0(\boldsymbol{r}, -\hat{\boldsymbol{k}}, -i\omega_n) =
a_0(\boldsymbol{r}, \hat{\boldsymbol{k}}, i\omega_n),\quad
\boldsymbol{a}
(\boldsymbol{r}, -\hat{\boldsymbol{k}}, - i\omega_n) = - \boldsymbol{a}
(\boldsymbol{r}, \hat{\boldsymbol{k}},  i\omega_n).
\end{align}
The spin-singlet component $a_0$ is either the even-parity even-frequency symmetry or
odd-party odd-frequency one.
The three spin-triplet component $\boldsymbol{a}$ are either the odd-parity even-frequency symmetry or
even-party odd-frequency one.
The Riccati parameter $\hat{a}$ in Eq.~(\ref{a_def_matsubara}) obeys the Eilenberger equation,
\begin{align}
i\hbar v_F &\hat{\boldsymbol{k}} \cdot \nabla \hat{a} + 2 i \omega_n \, \hat{a}
 + \boldsymbol{V}(\boldsymbol{r}) \cdot \boldsymbol{\sigma} \, \hat{a}
+  \hat{a} \, \boldsymbol{V}(\boldsymbol{r}) \cdot \boldsymbol{\sigma}
-i \;\mathrm{sgn}(\omega_n)\, \Delta(\boldsymbol{r})
+i \; \mathrm{sgn}(\omega_n)\, \hat{a}\, \underline{\Delta}(\boldsymbol{r}) \,
\hat{a}
=0, \label{eq:eilen_n1}
\end{align}
\begin{align}
i\hbar v_F &\hat{\boldsymbol{k}} \cdot \nabla \hat{\underline{a}} - 2 i \omega_n \, \hat{\underline{a}}
 - \boldsymbol{V}(\boldsymbol{r}) \cdot \boldsymbol{\sigma} \, \hat{\underline{a}}
-  \hat{\underline{a}} \, \boldsymbol{V}(\boldsymbol{r}) \cdot \boldsymbol{\sigma}
+i \; \mathrm{sgn}(\omega_n)\, \underline{\Delta}(\boldsymbol{r})
-i \; \mathrm{sgn}(\omega_n)\, \hat{\underline{a}}\, \Delta(\boldsymbol{r})
\, \hat{\underline{a}}=0.
\end{align}
The equation
\begin{align}
\underline{X}(\boldsymbol{r}, \hat{\boldsymbol{k}}, i \omega_n)= \hat{\sigma}_2\,
\undertilde{X}(\boldsymbol{r}, \hat{\boldsymbol{k}}, i \omega_n)\,
\hat{\sigma}_2=\hat{\sigma}_2\,
X^\ast(\boldsymbol{r}, -\hat{\boldsymbol{k}}, i \omega_n)\,
\hat{\sigma}_2,
\end{align}
defines the relation among $X$, $\undertilde{X}$, and $\underline{X}$.

The anomalous Green's function is represented by
\begin{align}
\hat{\mathcal{F}}(\boldsymbol{r}, \hat{\boldsymbol{k}}, i\omega_n) = & 2\,  \hat{\mathcal{N}}(\boldsymbol{r}, \hat{\boldsymbol{k}}, i\omega_n) \, \hat{a}(\boldsymbol{r}, \hat{\boldsymbol{k}}, i\omega_n)
= \mathcal{F}_0(\boldsymbol{r}, \hat{\boldsymbol{k}}, i\omega_n) 
+  \boldsymbol{\mathcal{F}}(\boldsymbol{r}, \hat{\boldsymbol{k}}, i\omega_n) 
\cdot \hat{\boldsymbol{\sigma}}, \label{eq:def_fcomponent}\\
%
 \mathcal{F}_0(\boldsymbol{r}, \hat{\boldsymbol{k}}, i\omega_n) = & \frac{2}{Z_N}\left[
 a_0 + (a_0^2 - \boldsymbol{a}^2)\, \underline{a}_0 \right]_{(\boldsymbol{r}, \hat{\boldsymbol{k}}, i\omega_n)}, \\
 \boldsymbol{\mathcal{F}}(\boldsymbol{r}, \hat{\boldsymbol{k}}, i\omega_n) = & \frac{2}{Z_N}\left[
\boldsymbol{a} + (a_0^2 - \boldsymbol{a}^2) \, \underline{\boldsymbol{a}} \right]_{(\boldsymbol{r}, \hat{\boldsymbol{k}}, i\omega_n)},\\
Z_N=& 1 + (a_0^2-\boldsymbol{a} \cdot \boldsymbol{a})(\underline{a}_0^2 -\underline{\boldsymbol{a}} \cdot \underline{\boldsymbol{a}})
+2 ( a_0 \, \underline{a}_0 - \boldsymbol{a} \cdot \underline{\boldsymbol{a}} ). \label{eq:zn_eilen}
\end{align}
The pair potential for spin-singlet $s$-wave symmetry 
is calculated self-consistently by using the gap equation,
\begin{align}
\Delta(\boldsymbol{r}) =& \pi \, N_0\, g \int \frac{d\hat{\boldsymbol{k}}}{S_d} 
T\sum_{\omega_n} \mathcal{F}_0(\boldsymbol{r}, \hat{\boldsymbol{k}}, i\omega_n), \label{eq:del_scf}
\end{align}
where $s_d$ is the solid angle in $d$-dimension and $g$ is the coupling constant.
The Ricatti parameters in the real energy representation $i\omega_n \to \epsilon + i\delta$ 
enable us to calculate the local density of states 
\begin{align}
N(\boldsymbol{r}, \epsilon)=&  N_0 \, \mathrm{Re} \int \frac{d\hat{\boldsymbol{k}}}{S_d} 
\left[ 2 \, \mathcal{N}_0(\boldsymbol{r}, \hat{\boldsymbol{k}}, \epsilon) -1 \right],\quad
\mathcal{\mathcal{N}}_0=\frac{1}{Z_N}\left( 1+ a_0\,  \underline{a}_0 - \boldsymbol{a} \cdot \underline{\boldsymbol{a}} \right),
\label{eq:ldos}
\end{align}
where $\mathcal{N}_0$ is the spin-singlet component of $\hat{\mathcal{N}}$ in Eq.~(\ref{n_def}).
The condensation energy of the superconducting states can be 
represented as~\cite{eilenberger:zphys1966},
\begin{align}
F_{\mathrm{S}}-F_{\mathrm{N}} = &\int d\boldsymbol{r} \, f_{\mathrm{SN}}(\boldsymbol{r}), \\
f_{\mathrm{SN}}(\boldsymbol{r}) = & 
{\pi N_0}  \, \int \frac{d\hat{\boldsymbol{k}}}{S_d} 
\left[ T\sum_{\omega_n} \Delta^\ast(\boldsymbol{r})\, 
\mathcal{F}_0(\boldsymbol{r}, \hat{\boldsymbol{k}}, i\omega_n) 
+ 8 T\sum_{\omega_n>0} \,\int_{\omega_n}^\infty d\omega 
\mathrm{Re}\left[ \mathcal{N}_0(\boldsymbol{r}, \hat{\boldsymbol{k}}, i\omega) -1 \right]
\right]. \label{eq:fsn}
\end{align}
The first term is derived from the constant term in the mean-field Hamiltonian.
Although the second term is calculated from the normal Green's function, 
 a Cooper pair affects the condensation energy through the normalization 
 of the Green's function, 
 (i.e, $\hat{\mathcal{G}}^2+\hat{\mathcal{F}}\underline{\hat{\mathcal{F}}}=\hat{1}$.) 


To discuss the suppression of $T_c$ due to odd-frequency Cooper pairs,
 we analyze the linearized Eilenberger equation.
By deleting the last term in Eq.(\ref{eq:eilen_n1}), we obtain 
\begin{align}
&( v_F \hat{\boldsymbol{k}} \cdot \nabla + 2 \omega_n) a_0 - 2i \boldsymbol{a} \cdot 
\boldsymbol{V}(\boldsymbol{r})
- \mathrm{sgn}(\omega_n) \,\Delta(\boldsymbol{r})=0, \label{eq:le1}\\
&( v_F \hat{\boldsymbol{k}} \cdot \nabla + 2 \omega_n) \boldsymbol{a} - 2i {a}_0 
 \boldsymbol{V}(\boldsymbol{r})=0.\label{eq:le2}
\end{align}
The first (second) equation corresponds to the spin-singlet (spin-triplet) part of 
 Eq.(\ref{eq:eilen_n1}).
The gap equation in the linear regime is given by,
\begin{align}
\Delta(\boldsymbol{r}) = \pi \, g\, N_0 \, T \sum_{\omega_n} \int \frac{d\hat{\boldsymbol{k}}}{S_d}\, 2
a_0 (\boldsymbol{r}, \hat{\boldsymbol{k}}, \omega_n).
\label{eq:lg1}
\end{align}

In numerical simulation in this paper, we calculate the Ricatti parameters $a(x, \pm, i\omega_n)$, 
where $+(-)$ indicates the positive (negative) momentum point on the Fermi surface in one-dimension.
The angle average on the Fermi surface is represented as
\begin{align}
\int \frac{d\hat{\boldsymbol{k}}}{S_d}  X(\boldsymbol{r}, \hat{\boldsymbol{k}}, i\omega_n)
\to \frac{1}{2}\sum_{s=\pm} X(x, s, i\omega_n).
\end{align} 

\end{widetext}
%


\begin{thebibliography}{54}%
\makeatletter
\providecommand \@ifxundefined [1]{%
 \@ifx{#1\undefined}
}%
\providecommand \@ifnum [1]{%
 \ifnum #1\expandafter \@firstoftwo
 \else \expandafter \@secondoftwo
 \fi
}%
\providecommand \@ifx [1]{%
 \ifx #1\expandafter \@firstoftwo
 \else \expandafter \@secondoftwo
 \fi
}%
\providecommand \natexlab [1]{#1}%
\providecommand \enquote  [1]{``#1''}%
\providecommand \bibnamefont  [1]{#1}%
\providecommand \bibfnamefont [1]{#1}%
\providecommand \citenamefont [1]{#1}%
\providecommand \href@noop [0]{\@secondoftwo}%
\providecommand \href [0]{\begingroup \@sanitize@url \@href}%
\providecommand \@href[1]{\@@startlink{#1}\@@href}%
\providecommand \@@href[1]{\endgroup#1\@@endlink}%
\providecommand \@sanitize@url [0]{\catcode `\\12\catcode `\$12\catcode
  `\&12\catcode `\#12\catcode `\^12\catcode `\_12\catcode `\%12\relax}%
\providecommand \@@startlink[1]{}%
\providecommand \@@endlink[0]{}%
\providecommand \url  [0]{\begingroup\@sanitize@url \@url }%
\providecommand \@url [1]{\endgroup\@href {#1}{\urlprefix }}%
\providecommand \urlprefix  [0]{URL }%
\providecommand \Eprint [0]{\href }%
\providecommand \doibase [0]{https://doi.org/}%
\providecommand \selectlanguage [0]{\@gobble}%
\providecommand \bibinfo  [0]{\@secondoftwo}%
\providecommand \bibfield  [0]{\@secondoftwo}%
\providecommand \translation [1]{[#1]}%
\providecommand \BibitemOpen [0]{}%
\providecommand \bibitemStop [0]{}%
\providecommand \bibitemNoStop [0]{.\EOS\space}%
\providecommand \EOS [0]{\spacefactor3000\relax}%
\providecommand \BibitemShut  [1]{\csname bibitem#1\endcsname}%
\let\auto@bib@innerbib\@empty
\bibitem [{\citenamefont {Yu}(1965)}]{yu:actphys1965}%
  \BibitemOpen
  \bibfield  {author} {\bibinfo {author} {\bibfnamefont {L.}~\bibnamefont
  {Yu}},\ }\bibfield  {title} {\bibinfo {title} {Bound state in superconductors
  with paramagnetic impurities},\ }\href@noop {} {\bibfield  {journal}
  {\bibinfo  {journal} {Acta. Phys. Sin}\ }\textbf {\bibinfo {volume} {21}},\
  \bibinfo {pages} {75} (\bibinfo {year} {1965})}\BibitemShut {NoStop}%
\bibitem [{\citenamefont {Shiba}(1968)}]{shiba:ptp1968}%
  \BibitemOpen
  \bibfield  {author} {\bibinfo {author} {\bibfnamefont {H.}~\bibnamefont
  {Shiba}},\ }\bibfield  {title} {\bibinfo {title} {{Classical Spins in
  Superconductors}},\ }\href {https://doi.org/10.1143/PTP.40.435} {\bibfield
  {journal} {\bibinfo  {journal} {Progress of Theoretical Physics}\ }\textbf
  {\bibinfo {volume} {40}},\ \bibinfo {pages} {435} (\bibinfo {year} {1968})},\
  \Eprint
  {https://arxiv.org/abs/https://academic.oup.com/ptp/article-pdf/40/3/435/5185550/40-3-435.pdf}
  {https://academic.oup.com/ptp/article-pdf/40/3/435/5185550/40-3-435.pdf}
  \BibitemShut {NoStop}%
\bibitem [{\citenamefont {Rusinov}(1969)}]{rusinov:jetplett1969}%
  \BibitemOpen
  \bibfield  {author} {\bibinfo {author} {\bibfnamefont {A.~I.}\ \bibnamefont
  {Rusinov}},\ }\bibfield  {title} {\bibinfo {title} {Superconductivity near a
  paramagnetic impurity},\ }\href@noop {} {\bibfield  {journal} {\bibinfo
  {journal} {Sov. Phys. JETP Lett}\ }\textbf {\bibinfo {volume} {9}},\ \bibinfo
  {pages} {85} (\bibinfo {year} {1969})}\BibitemShut {NoStop}%
\bibitem [{\citenamefont {Choy}\ \emph {et~al.}(2011)\citenamefont {Choy},
  \citenamefont {Edge}, \citenamefont {Akhmerov},\ and\ \citenamefont
  {Beenakker}}]{choy:prb2011}%
  \BibitemOpen
  \bibfield  {author} {\bibinfo {author} {\bibfnamefont {T.-P.}\ \bibnamefont
  {Choy}}, \bibinfo {author} {\bibfnamefont {J.~M.}\ \bibnamefont {Edge}},
  \bibinfo {author} {\bibfnamefont {A.~R.}\ \bibnamefont {Akhmerov}},\ and\
  \bibinfo {author} {\bibfnamefont {C.~W.~J.}\ \bibnamefont {Beenakker}},\
  }\bibfield  {title} {\bibinfo {title} {Majorana fermions emerging from
  magnetic nanoparticles on a superconductor without spin-orbit coupling},\
  }\href {https://doi.org/10.1103/PhysRevB.84.195442} {\bibfield  {journal}
  {\bibinfo  {journal} {Phys. Rev. B}\ }\textbf {\bibinfo {volume} {84}},\
  \bibinfo {pages} {195442} (\bibinfo {year} {2011})}\BibitemShut {NoStop}%
\bibitem [{\citenamefont {Klinovaja}\ \emph {et~al.}(2013)\citenamefont
  {Klinovaja}, \citenamefont {Stano}, \citenamefont {Yazdani},\ and\
  \citenamefont {Loss}}]{klinovaja:prl2013}%
  \BibitemOpen
  \bibfield  {author} {\bibinfo {author} {\bibfnamefont {J.}~\bibnamefont
  {Klinovaja}}, \bibinfo {author} {\bibfnamefont {P.}~\bibnamefont {Stano}},
  \bibinfo {author} {\bibfnamefont {A.}~\bibnamefont {Yazdani}},\ and\ \bibinfo
  {author} {\bibfnamefont {D.}~\bibnamefont {Loss}},\ }\bibfield  {title}
  {\bibinfo {title} {Topological superconductivity and majorana fermions in
  rkky systems},\ }\href {https://doi.org/10.1103/PhysRevLett.111.186805}
  {\bibfield  {journal} {\bibinfo  {journal} {Phys. Rev. Lett.}\ }\textbf
  {\bibinfo {volume} {111}},\ \bibinfo {pages} {186805} (\bibinfo {year}
  {2013})}\BibitemShut {NoStop}%
\bibitem [{\citenamefont {Nadj-Perge}\ \emph {et~al.}(2014)\citenamefont
  {Nadj-Perge}, \citenamefont {Drozdov}, \citenamefont {Li}, \citenamefont
  {Chen}, \citenamefont {Jeon}, \citenamefont {Seo}, \citenamefont {MacDonald},
  \citenamefont {Bernevig},\ and\ \citenamefont
  {Yazdani}}]{Nadjerge:science2014}%
  \BibitemOpen
  \bibfield  {author} {\bibinfo {author} {\bibfnamefont {S.}~\bibnamefont
  {Nadj-Perge}}, \bibinfo {author} {\bibfnamefont {I.~K.}\ \bibnamefont
  {Drozdov}}, \bibinfo {author} {\bibfnamefont {J.}~\bibnamefont {Li}},
  \bibinfo {author} {\bibfnamefont {H.}~\bibnamefont {Chen}}, \bibinfo {author}
  {\bibfnamefont {S.}~\bibnamefont {Jeon}}, \bibinfo {author} {\bibfnamefont
  {J.}~\bibnamefont {Seo}}, \bibinfo {author} {\bibfnamefont {A.~H.}\
  \bibnamefont {MacDonald}}, \bibinfo {author} {\bibfnamefont {B.~A.}\
  \bibnamefont {Bernevig}},\ and\ \bibinfo {author} {\bibfnamefont
  {A.}~\bibnamefont {Yazdani}},\ }\bibfield  {title} {\bibinfo {title}
  {Observation of majorana fermions in ferromagnetic atomic chains on a
  superconductor},\ }\href {https://doi.org/10.1126/science.1259327} {\bibfield
   {journal} {\bibinfo  {journal} {Science}\ }\textbf {\bibinfo {volume}
  {346}},\ \bibinfo {pages} {602} (\bibinfo {year} {2014})},\ \Eprint
  {https://arxiv.org/abs/https://www.science.org/doi/pdf/10.1126/science.1259327}
  {https://www.science.org/doi/pdf/10.1126/science.1259327} \BibitemShut
  {NoStop}%
\bibitem [{\citenamefont {M{\'e}nard}\ \emph {et~al.}(2015)\citenamefont
  {M{\'e}nard}, \citenamefont {Guissart}, \citenamefont {Brun}, \citenamefont
  {Pons}, \citenamefont {Stolyarov}, \citenamefont {Debontridder},
  \citenamefont {Leclerc}, \citenamefont {Janod}, \citenamefont {Cario},
  \citenamefont {Roditchev}, \citenamefont {Simon},\ and\ \citenamefont
  {Cren}}]{menard:natphys2015}%
  \BibitemOpen
  \bibfield  {author} {\bibinfo {author} {\bibfnamefont {G.~C.}\ \bibnamefont
  {M{\'e}nard}}, \bibinfo {author} {\bibfnamefont {S.}~\bibnamefont
  {Guissart}}, \bibinfo {author} {\bibfnamefont {C.}~\bibnamefont {Brun}},
  \bibinfo {author} {\bibfnamefont {S.}~\bibnamefont {Pons}}, \bibinfo {author}
  {\bibfnamefont {V.~S.}\ \bibnamefont {Stolyarov}}, \bibinfo {author}
  {\bibfnamefont {F.}~\bibnamefont {Debontridder}}, \bibinfo {author}
  {\bibfnamefont {M.~V.}\ \bibnamefont {Leclerc}}, \bibinfo {author}
  {\bibfnamefont {E.}~\bibnamefont {Janod}}, \bibinfo {author} {\bibfnamefont
  {L.}~\bibnamefont {Cario}}, \bibinfo {author} {\bibfnamefont
  {D.}~\bibnamefont {Roditchev}}, \bibinfo {author} {\bibfnamefont
  {P.}~\bibnamefont {Simon}},\ and\ \bibinfo {author} {\bibfnamefont
  {T.}~\bibnamefont {Cren}},\ }\bibfield  {title} {\bibinfo {title} {Coherent
  long-range magnetic bound states in a^^c2^^a0superconductor},\ }\href
  {https://doi.org/10.1038/nphys3508} {\bibfield  {journal} {\bibinfo
  {journal} {Nature Physics}\ }\textbf {\bibinfo {volume} {11}},\ \bibinfo
  {pages} {1013} (\bibinfo {year} {2015})}\BibitemShut {NoStop}%
\bibitem [{\citenamefont {Scher{\"u}bl}\ \emph {et~al.}(2020)\citenamefont
  {Scher{\"u}bl}, \citenamefont {F{\"u}l{\"o}p}, \citenamefont {Moca},
  \citenamefont {Gramich}, \citenamefont {Baumgartner}, \citenamefont {Makk},
  \citenamefont {Elalaily}, \citenamefont {Sch{\"o}nenberger}, \citenamefont
  {Nyg{\aa}rd}, \citenamefont {Zar{\'a}nd},\ and\ \citenamefont
  {Csonka}}]{scherubl:natphys2020}%
  \BibitemOpen
  \bibfield  {author} {\bibinfo {author} {\bibfnamefont {Z.}~\bibnamefont
  {Scher{\"u}bl}}, \bibinfo {author} {\bibfnamefont {G.}~\bibnamefont
  {F{\"u}l{\"o}p}}, \bibinfo {author} {\bibfnamefont {C.~P.}\ \bibnamefont
  {Moca}}, \bibinfo {author} {\bibfnamefont {J.}~\bibnamefont {Gramich}},
  \bibinfo {author} {\bibfnamefont {A.}~\bibnamefont {Baumgartner}}, \bibinfo
  {author} {\bibfnamefont {P.}~\bibnamefont {Makk}}, \bibinfo {author}
  {\bibfnamefont {T.}~\bibnamefont {Elalaily}}, \bibinfo {author}
  {\bibfnamefont {C.}~\bibnamefont {Sch{\"o}nenberger}}, \bibinfo {author}
  {\bibfnamefont {J.}~\bibnamefont {Nyg{\aa}rd}}, \bibinfo {author}
  {\bibfnamefont {G.}~\bibnamefont {Zar{\'a}nd}},\ and\ \bibinfo {author}
  {\bibfnamefont {S.}~\bibnamefont {Csonka}},\ }\bibfield  {title} {\bibinfo
  {title} {Large spatial extension of the zero-energy yu--shiba--rusinov state
  in a magnetic field},\ }\href {https://doi.org/10.1038/s41467-020-15322-9}
  {\bibfield  {journal} {\bibinfo  {journal} {Nature Communications}\ }\textbf
  {\bibinfo {volume} {11}},\ \bibinfo {pages} {1834} (\bibinfo {year}
  {2020})}\BibitemShut {NoStop}%
\bibitem [{\citenamefont {Ebisu}\ \emph {et~al.}(2015)\citenamefont {Ebisu},
  \citenamefont {Yada}, \citenamefont {Kasai},\ and\ \citenamefont
  {Tanaka}}]{ebis:prb2015}%
  \BibitemOpen
  \bibfield  {author} {\bibinfo {author} {\bibfnamefont {H.}~\bibnamefont
  {Ebisu}}, \bibinfo {author} {\bibfnamefont {K.}~\bibnamefont {Yada}},
  \bibinfo {author} {\bibfnamefont {H.}~\bibnamefont {Kasai}},\ and\ \bibinfo
  {author} {\bibfnamefont {Y.}~\bibnamefont {Tanaka}},\ }\bibfield  {title}
  {\bibinfo {title} {Odd-frequency pairing in topological superconductivity in
  a one-dimensional magnetic chain},\ }\href
  {https://doi.org/10.1103/PhysRevB.91.054518} {\bibfield  {journal} {\bibinfo
  {journal} {Phys. Rev. B}\ }\textbf {\bibinfo {volume} {91}},\ \bibinfo
  {pages} {054518} (\bibinfo {year} {2015})}\BibitemShut {NoStop}%
\bibitem [{\citenamefont {Abrikosov}\ and\ \citenamefont
  {Gor{\textquoteright}kov}(1961)}]{abrikosov:jetp1961}%
  \BibitemOpen
  \bibfield  {author} {\bibinfo {author} {\bibfnamefont {A.~A.}\ \bibnamefont
  {Abrikosov}}\ and\ \bibinfo {author} {\bibfnamefont {L.~P.}\ \bibnamefont
  {Gor{\textquoteright}kov}},\ }\bibfield  {title} {\bibinfo {title}
  {Contribution to the theory of superconducting alloys with paramagnetic
  impurities},\ }\href@noop {} {\bibfield  {journal} {\bibinfo  {journal} {Sov.
  Phys. JETP}\ }\textbf {\bibinfo {volume} {12}},\ \bibinfo {pages} {1243}
  (\bibinfo {year} {1961})}\BibitemShut {NoStop}%
\bibitem [{\citenamefont {Maki}(1969)}]{maki:book1969}%
  \BibitemOpen
  \bibfield  {author} {\bibinfo {author} {\bibfnamefont {K.}~\bibnamefont
  {Maki}},\ }\href@noop {} {\emph {\bibinfo {title} {Superconductivity}}},\
  edited by\ \bibinfo {editor} {\bibfnamefont {R.~D.}\ \bibnamefont {Parks}},\
  Vol.~\bibinfo {volume} {2}\ (\bibinfo  {publisher} {Marcel Dekker},\ \bibinfo
  {year} {1969})\ p.\ \bibinfo {pages} {1035}\BibitemShut {NoStop}%
\bibitem [{\citenamefont {Salkola}\ \emph {et~al.}(1997)\citenamefont
  {Salkola}, \citenamefont {Balatsky},\ and\ \citenamefont
  {Schrieffer}}]{salkola:prb1997}%
  \BibitemOpen
  \bibfield  {author} {\bibinfo {author} {\bibfnamefont {M.~I.}\ \bibnamefont
  {Salkola}}, \bibinfo {author} {\bibfnamefont {A.~V.}\ \bibnamefont
  {Balatsky}},\ and\ \bibinfo {author} {\bibfnamefont {J.~R.}\ \bibnamefont
  {Schrieffer}},\ }\bibfield  {title} {\bibinfo {title} {Spectral properties of
  quasiparticle excitations induced by magnetic moments in superconductors},\
  }\href {https://doi.org/10.1103/PhysRevB.55.12648} {\bibfield  {journal}
  {\bibinfo  {journal} {Phys. Rev. B}\ }\textbf {\bibinfo {volume} {55}},\
  \bibinfo {pages} {12648} (\bibinfo {year} {1997})}\BibitemShut {NoStop}%
\bibitem [{\citenamefont {Flatt\'e}\ and\ \citenamefont
  {Byers}(1997)}]{flatte:prl1997}%
  \BibitemOpen
  \bibfield  {author} {\bibinfo {author} {\bibfnamefont {M.~E.}\ \bibnamefont
  {Flatt\'e}}\ and\ \bibinfo {author} {\bibfnamefont {J.~M.}\ \bibnamefont
  {Byers}},\ }\bibfield  {title} {\bibinfo {title} {Local electronic structure
  of a single magnetic impurity in a superconductor},\ }\href
  {https://doi.org/10.1103/PhysRevLett.78.3761} {\bibfield  {journal} {\bibinfo
   {journal} {Phys. Rev. Lett.}\ }\textbf {\bibinfo {volume} {78}},\ \bibinfo
  {pages} {3761} (\bibinfo {year} {1997})}\BibitemShut {NoStop}%
\bibitem [{\citenamefont {Meng}\ \emph {et~al.}(2015)\citenamefont {Meng},
  \citenamefont {Klinovaja}, \citenamefont {Hoffman}, \citenamefont {Simon},\
  and\ \citenamefont {Loss}}]{meng:prb2015}%
  \BibitemOpen
  \bibfield  {author} {\bibinfo {author} {\bibfnamefont {T.}~\bibnamefont
  {Meng}}, \bibinfo {author} {\bibfnamefont {J.}~\bibnamefont {Klinovaja}},
  \bibinfo {author} {\bibfnamefont {S.}~\bibnamefont {Hoffman}}, \bibinfo
  {author} {\bibfnamefont {P.}~\bibnamefont {Simon}},\ and\ \bibinfo {author}
  {\bibfnamefont {D.}~\bibnamefont {Loss}},\ }\bibfield  {title} {\bibinfo
  {title} {Superconducting gap renormalization around two magnetic impurities:
  From shiba to andreev bound states},\ }\href
  {https://doi.org/10.1103/PhysRevB.92.064503} {\bibfield  {journal} {\bibinfo
  {journal} {Phys. Rev. B}\ }\textbf {\bibinfo {volume} {92}},\ \bibinfo
  {pages} {064503} (\bibinfo {year} {2015})}\BibitemShut {NoStop}%
\bibitem [{\citenamefont {Bj\"ornson}\ \emph {et~al.}(2017)\citenamefont
  {Bj\"ornson}, \citenamefont {Balatsky},\ and\ \citenamefont
  {Black-Schaffer}}]{bjornson:prb2017}%
  \BibitemOpen
  \bibfield  {author} {\bibinfo {author} {\bibfnamefont {K.}~\bibnamefont
  {Bj\"ornson}}, \bibinfo {author} {\bibfnamefont {A.~V.}\ \bibnamefont
  {Balatsky}},\ and\ \bibinfo {author} {\bibfnamefont {A.~M.}\ \bibnamefont
  {Black-Schaffer}},\ }\bibfield  {title} {\bibinfo {title} {Superconducting
  order parameter $\ensuremath{\pi}$-phase shift in magnetic impurity wires},\
  }\href {https://doi.org/10.1103/PhysRevB.95.104521} {\bibfield  {journal}
  {\bibinfo  {journal} {Phys. Rev. B}\ }\textbf {\bibinfo {volume} {95}},\
  \bibinfo {pages} {104521} (\bibinfo {year} {2017})}\BibitemShut {NoStop}%
\bibitem [{\citenamefont {Balatsky}\ \emph {et~al.}(2006)\citenamefont
  {Balatsky}, \citenamefont {Vekhter},\ and\ \citenamefont
  {Zhu}}]{balatsky:rmp2006}%
  \BibitemOpen
  \bibfield  {author} {\bibinfo {author} {\bibfnamefont {A.~V.}\ \bibnamefont
  {Balatsky}}, \bibinfo {author} {\bibfnamefont {I.}~\bibnamefont {Vekhter}},\
  and\ \bibinfo {author} {\bibfnamefont {J.-X.}\ \bibnamefont {Zhu}},\
  }\bibfield  {title} {\bibinfo {title} {Impurity-induced states in
  conventional and unconventional superconductors},\ }\href
  {https://doi.org/10.1103/RevModPhys.78.373} {\bibfield  {journal} {\bibinfo
  {journal} {Rev. Mod. Phys.}\ }\textbf {\bibinfo {volume} {78}},\ \bibinfo
  {pages} {373} (\bibinfo {year} {2006})}\BibitemShut {NoStop}%
\bibitem [{\citenamefont {Berezinskii}(1974)}]{berezinskii:jetplett1974}%
  \BibitemOpen
  \bibfield  {author} {\bibinfo {author} {\bibfnamefont {V.~L.}\ \bibnamefont
  {Berezinskii}},\ }\bibfield  {title} {\bibinfo {title} {New model of the
  anisotropic phase of superfluid he$^3$},\ }\href@noop {} {\bibfield
  {journal} {\bibinfo  {journal} {JETP Lett.}\ }\textbf {\bibinfo {volume}
  {20}},\ \bibinfo {pages} {287} (\bibinfo {year} {1974})}\BibitemShut
  {NoStop}%
\bibitem [{\citenamefont {Kirkpatrick}\ and\ \citenamefont
  {Belitz}(1991)}]{kirkpatrik:prl1991}%
  \BibitemOpen
  \bibfield  {author} {\bibinfo {author} {\bibfnamefont {T.~R.}\ \bibnamefont
  {Kirkpatrick}}\ and\ \bibinfo {author} {\bibfnamefont {D.}~\bibnamefont
  {Belitz}},\ }\bibfield  {title} {\bibinfo {title} {Disorder-induced triplet
  superconductivity},\ }\href {https://doi.org/10.1103/PhysRevLett.66.1533}
  {\bibfield  {journal} {\bibinfo  {journal} {Phys. Rev. Lett.}\ }\textbf
  {\bibinfo {volume} {66}},\ \bibinfo {pages} {1533} (\bibinfo {year}
  {1991})}\BibitemShut {NoStop}%
\bibitem [{\citenamefont {Belitz}\ and\ \citenamefont
  {Kirkpatrick}(1992)}]{belitz:prb1992}%
  \BibitemOpen
  \bibfield  {author} {\bibinfo {author} {\bibfnamefont {D.}~\bibnamefont
  {Belitz}}\ and\ \bibinfo {author} {\bibfnamefont {T.~R.}\ \bibnamefont
  {Kirkpatrick}},\ }\bibfield  {title} {\bibinfo {title} {Even-parity
  spin-triplet superconductivity in disordered electronic systems},\ }\href
  {https://doi.org/10.1103/PhysRevB.46.8393} {\bibfield  {journal} {\bibinfo
  {journal} {Phys. Rev. B}\ }\textbf {\bibinfo {volume} {46}},\ \bibinfo
  {pages} {8393} (\bibinfo {year} {1992})}\BibitemShut {NoStop}%
\bibitem [{\citenamefont {Balatsky}\ and\ \citenamefont
  {Abrahams}(1992)}]{balatsky:prb1992}%
  \BibitemOpen
  \bibfield  {author} {\bibinfo {author} {\bibfnamefont {A.}~\bibnamefont
  {Balatsky}}\ and\ \bibinfo {author} {\bibfnamefont {E.}~\bibnamefont
  {Abrahams}},\ }\bibfield  {title} {\bibinfo {title} {New class of singlet
  superconductors which break the time reversal and parity},\ }\href
  {https://doi.org/10.1103/PhysRevB.45.13125} {\bibfield  {journal} {\bibinfo
  {journal} {Phys. Rev. B}\ }\textbf {\bibinfo {volume} {45}},\ \bibinfo
  {pages} {13125} (\bibinfo {year} {1992})}\BibitemShut {NoStop}%
\bibitem [{\citenamefont {Abrahams}\ \emph {et~al.}(1993)\citenamefont
  {Abrahams}, \citenamefont {Balatsky}, \citenamefont {Schrieffer},\ and\
  \citenamefont {Allen}}]{abrahams:prb1993}%
  \BibitemOpen
  \bibfield  {author} {\bibinfo {author} {\bibfnamefont {E.}~\bibnamefont
  {Abrahams}}, \bibinfo {author} {\bibfnamefont {A.}~\bibnamefont {Balatsky}},
  \bibinfo {author} {\bibfnamefont {J.~R.}\ \bibnamefont {Schrieffer}},\ and\
  \bibinfo {author} {\bibfnamefont {P.~B.}\ \bibnamefont {Allen}},\ }\bibfield
  {title} {\bibinfo {title} {Interactions for odd-\ensuremath{\omega}-gap
  singlet superconductors},\ }\href {https://doi.org/10.1103/PhysRevB.47.513}
  {\bibfield  {journal} {\bibinfo  {journal} {Phys. Rev. B}\ }\textbf {\bibinfo
  {volume} {47}},\ \bibinfo {pages} {513} (\bibinfo {year} {1993})}\BibitemShut
  {NoStop}%
\bibitem [{\citenamefont {Coleman}\ \emph {et~al.}(1993)\citenamefont
  {Coleman}, \citenamefont {Miranda},\ and\ \citenamefont
  {Tsvelik}}]{coleman:prl1993}%
  \BibitemOpen
  \bibfield  {author} {\bibinfo {author} {\bibfnamefont {P.}~\bibnamefont
  {Coleman}}, \bibinfo {author} {\bibfnamefont {E.}~\bibnamefont {Miranda}},\
  and\ \bibinfo {author} {\bibfnamefont {A.}~\bibnamefont {Tsvelik}},\
  }\bibfield  {title} {\bibinfo {title} {Possible realization of odd-frequency
  pairing in heavy fermion compounds},\ }\href
  {https://doi.org/10.1103/PhysRevLett.70.2960} {\bibfield  {journal} {\bibinfo
   {journal} {Phys. Rev. Lett.}\ }\textbf {\bibinfo {volume} {70}},\ \bibinfo
  {pages} {2960} (\bibinfo {year} {1993})}\BibitemShut {NoStop}%
\bibitem [{\citenamefont {Abrahams}\ \emph {et~al.}(1995)\citenamefont
  {Abrahams}, \citenamefont {Balatsky}, \citenamefont {Scalapino},\ and\
  \citenamefont {Schrieffer}}]{abrahams:prb1995}%
  \BibitemOpen
  \bibfield  {author} {\bibinfo {author} {\bibfnamefont {E.}~\bibnamefont
  {Abrahams}}, \bibinfo {author} {\bibfnamefont {A.}~\bibnamefont {Balatsky}},
  \bibinfo {author} {\bibfnamefont {D.~J.}\ \bibnamefont {Scalapino}},\ and\
  \bibinfo {author} {\bibfnamefont {J.~R.}\ \bibnamefont {Schrieffer}},\
  }\bibfield  {title} {\bibinfo {title} {Properties of odd-gap
  superconductors},\ }\href {https://doi.org/10.1103/PhysRevB.52.1271}
  {\bibfield  {journal} {\bibinfo  {journal} {Phys. Rev. B}\ }\textbf {\bibinfo
  {volume} {52}},\ \bibinfo {pages} {1271} (\bibinfo {year}
  {1995})}\BibitemShut {NoStop}%
\bibitem [{\citenamefont {Zachar}\ \emph {et~al.}(1996)\citenamefont {Zachar},
  \citenamefont {Kivelson},\ and\ \citenamefont {Emery}}]{zachar:prl1996}%
  \BibitemOpen
  \bibfield  {author} {\bibinfo {author} {\bibfnamefont {O.}~\bibnamefont
  {Zachar}}, \bibinfo {author} {\bibfnamefont {S.~A.}\ \bibnamefont
  {Kivelson}},\ and\ \bibinfo {author} {\bibfnamefont {V.~J.}\ \bibnamefont
  {Emery}},\ }\bibfield  {title} {\bibinfo {title} {Exact results for a 1d
  kondo lattice from bosonization},\ }\href
  {https://doi.org/10.1103/PhysRevLett.77.1342} {\bibfield  {journal} {\bibinfo
   {journal} {Phys. Rev. Lett.}\ }\textbf {\bibinfo {volume} {77}},\ \bibinfo
  {pages} {1342} (\bibinfo {year} {1996})}\BibitemShut {NoStop}%
\bibitem [{\citenamefont {Vojta}\ and\ \citenamefont
  {Dagotto}(1999)}]{vojta:prb1999}%
  \BibitemOpen
  \bibfield  {author} {\bibinfo {author} {\bibfnamefont {M.}~\bibnamefont
  {Vojta}}\ and\ \bibinfo {author} {\bibfnamefont {E.}~\bibnamefont
  {Dagotto}},\ }\bibfield  {title} {\bibinfo {title} {Indications of
  unconventional superconductivity in doped and undoped triangular
  antiferromagnets},\ }\href {https://doi.org/10.1103/PhysRevB.59.R713}
  {\bibfield  {journal} {\bibinfo  {journal} {Phys. Rev. B}\ }\textbf {\bibinfo
  {volume} {59}},\ \bibinfo {pages} {R713} (\bibinfo {year}
  {1999})}\BibitemShut {NoStop}%
\bibitem [{\citenamefont {Fominov}\ \emph {et~al.}(2015)\citenamefont
  {Fominov}, \citenamefont {Tanaka}, \citenamefont {Asano},\ and\ \citenamefont
  {Eschrig}}]{fominov:prb2015}%
  \BibitemOpen
  \bibfield  {author} {\bibinfo {author} {\bibfnamefont {Y.~V.}\ \bibnamefont
  {Fominov}}, \bibinfo {author} {\bibfnamefont {Y.}~\bibnamefont {Tanaka}},
  \bibinfo {author} {\bibfnamefont {Y.}~\bibnamefont {Asano}},\ and\ \bibinfo
  {author} {\bibfnamefont {M.}~\bibnamefont {Eschrig}},\ }\bibfield  {title}
  {\bibinfo {title} {Odd-frequency superconducting states with different types
  of meissner response: Problem of coexistence},\ }\href
  {https://doi.org/10.1103/PhysRevB.91.144514} {\bibfield  {journal} {\bibinfo
  {journal} {Phys. Rev. B}\ }\textbf {\bibinfo {volume} {91}},\ \bibinfo
  {pages} {144514} (\bibinfo {year} {2015})}\BibitemShut {NoStop}%
\bibitem [{\citenamefont {Bergeret}\ \emph {et~al.}(2001)\citenamefont
  {Bergeret}, \citenamefont {Volkov},\ and\ \citenamefont
  {Efetov}}]{bergeret:prl2001}%
  \BibitemOpen
  \bibfield  {author} {\bibinfo {author} {\bibfnamefont {F.~S.}\ \bibnamefont
  {Bergeret}}, \bibinfo {author} {\bibfnamefont {A.~F.}\ \bibnamefont
  {Volkov}},\ and\ \bibinfo {author} {\bibfnamefont {K.~B.}\ \bibnamefont
  {Efetov}},\ }\bibfield  {title} {\bibinfo {title} {Long-range proximity
  effects in superconductor-ferromagnet structures},\ }\href
  {https://doi.org/10.1103/PhysRevLett.86.4096} {\bibfield  {journal} {\bibinfo
   {journal} {Phys. Rev. Lett.}\ }\textbf {\bibinfo {volume} {86}},\ \bibinfo
  {pages} {4096} (\bibinfo {year} {2001})}\BibitemShut {NoStop}%
\bibitem [{\citenamefont {Buchholtz}\ and\ \citenamefont
  {Zwicknagl}(1981)}]{buchholz:prb1981}%
  \BibitemOpen
  \bibfield  {author} {\bibinfo {author} {\bibfnamefont {L.~J.}\ \bibnamefont
  {Buchholtz}}\ and\ \bibinfo {author} {\bibfnamefont {G.}~\bibnamefont
  {Zwicknagl}},\ }\bibfield  {title} {\bibinfo {title} {Identification of
  $p$-wave superconductors},\ }\href {https://doi.org/10.1103/PhysRevB.23.5788}
  {\bibfield  {journal} {\bibinfo  {journal} {Phys. Rev. B}\ }\textbf {\bibinfo
  {volume} {23}},\ \bibinfo {pages} {5788} (\bibinfo {year}
  {1981})}\BibitemShut {NoStop}%
\bibitem [{\citenamefont {Hara}\ and\ \citenamefont
  {Nagai}(1986)}]{hara:ptp1986}%
  \BibitemOpen
  \bibfield  {author} {\bibinfo {author} {\bibfnamefont {J.}~\bibnamefont
  {Hara}}\ and\ \bibinfo {author} {\bibfnamefont {K.}~\bibnamefont {Nagai}},\
  }\bibfield  {title} {\bibinfo {title} {{A Polar State in a Slab as a Soluble
  Model of p-Wave Fermi Superfluid in Finite Geometry}},\ }\href
  {https://doi.org/10.1143/PTP.76.1237} {\bibfield  {journal} {\bibinfo
  {journal} {Progress of Theoretical Physics}\ }\textbf {\bibinfo {volume}
  {76}},\ \bibinfo {pages} {1237} (\bibinfo {year} {1986})},\ \Eprint
  {https://arxiv.org/abs/https://academic.oup.com/ptp/article-pdf/76/6/1237/5344780/76-6-1237.pdf}
  {https://academic.oup.com/ptp/article-pdf/76/6/1237/5344780/76-6-1237.pdf}
  \BibitemShut {NoStop}%
\bibitem [{\citenamefont {Hu}(1994)}]{hu:prl1994}%
  \BibitemOpen
  \bibfield  {author} {\bibinfo {author} {\bibfnamefont {C.-R.}\ \bibnamefont
  {Hu}},\ }\bibfield  {title} {\bibinfo {title} {Midgap surface states as a
  novel signature for
  ${\mathit{d}}_{\mathit{x}\mathit{a}}^{2}$-${\mathit{x}}_{\mathit{b}}^{2}$-wave
  superconductivity},\ }\href {https://doi.org/10.1103/PhysRevLett.72.1526}
  {\bibfield  {journal} {\bibinfo  {journal} {Phys. Rev. Lett.}\ }\textbf
  {\bibinfo {volume} {72}},\ \bibinfo {pages} {1526} (\bibinfo {year}
  {1994})}\BibitemShut {NoStop}%
\bibitem [{\citenamefont {Tanaka}\ and\ \citenamefont
  {Kashiwaya}(1995)}]{tanaka:prl1995}%
  \BibitemOpen
  \bibfield  {author} {\bibinfo {author} {\bibfnamefont {Y.}~\bibnamefont
  {Tanaka}}\ and\ \bibinfo {author} {\bibfnamefont {S.}~\bibnamefont
  {Kashiwaya}},\ }\bibfield  {title} {\bibinfo {title} {Theory of tunneling
  spectroscopy of $\mathit{d}$-wave superconductors},\ }\href
  {https://doi.org/10.1103/PhysRevLett.74.3451} {\bibfield  {journal} {\bibinfo
   {journal} {Phys. Rev. Lett.}\ }\textbf {\bibinfo {volume} {74}},\ \bibinfo
  {pages} {3451} (\bibinfo {year} {1995})}\BibitemShut {NoStop}%
\bibitem [{\citenamefont {Tanaka}\ \emph {et~al.}(2012)\citenamefont {Tanaka},
  \citenamefont {Sato},\ and\ \citenamefont {Nagaosa}}]{tanaka:jpsj2012}%
  \BibitemOpen
  \bibfield  {author} {\bibinfo {author} {\bibfnamefont {Y.}~\bibnamefont
  {Tanaka}}, \bibinfo {author} {\bibfnamefont {M.}~\bibnamefont {Sato}},\ and\
  \bibinfo {author} {\bibfnamefont {N.}~\bibnamefont {Nagaosa}},\ }\bibfield
  {title} {\bibinfo {title} {Symmetry and topology in superconductors
  ^^e2^^80^^93odd-frequency pairing and edge states^^e2^^80^^93},\ }\href
  {https://doi.org/10.1143/JPSJ.81.011013} {\bibfield  {journal} {\bibinfo
  {journal} {Journal of the Physical Society of Japan}\ }\textbf {\bibinfo
  {volume} {81}},\ \bibinfo {pages} {011013} (\bibinfo {year} {2012})},\
  \Eprint {https://arxiv.org/abs/https://doi.org/10.1143/JPSJ.81.011013}
  {https://doi.org/10.1143/JPSJ.81.011013} \BibitemShut {NoStop}%
\bibitem [{\citenamefont {Yokoyama}\ \emph {et~al.}(2008)\citenamefont
  {Yokoyama}, \citenamefont {Tanaka},\ and\ \citenamefont
  {Golubov}}]{yokoyama:prb2008}%
  \BibitemOpen
  \bibfield  {author} {\bibinfo {author} {\bibfnamefont {T.}~\bibnamefont
  {Yokoyama}}, \bibinfo {author} {\bibfnamefont {Y.}~\bibnamefont {Tanaka}},\
  and\ \bibinfo {author} {\bibfnamefont {A.~A.}\ \bibnamefont {Golubov}},\
  }\bibfield  {title} {\bibinfo {title} {Theory of pairing symmetry inside the
  abrikosov vortex core},\ }\href {https://doi.org/10.1103/PhysRevB.78.012508}
  {\bibfield  {journal} {\bibinfo  {journal} {Phys. Rev. B}\ }\textbf {\bibinfo
  {volume} {78}},\ \bibinfo {pages} {012508} (\bibinfo {year}
  {2008})}\BibitemShut {NoStop}%
\bibitem [{\citenamefont {Tanuma}\ \emph {et~al.}(2009)\citenamefont {Tanuma},
  \citenamefont {Hayashi}, \citenamefont {Tanaka},\ and\ \citenamefont
  {Golubov}}]{tanuma:prl2009}%
  \BibitemOpen
  \bibfield  {author} {\bibinfo {author} {\bibfnamefont {Y.}~\bibnamefont
  {Tanuma}}, \bibinfo {author} {\bibfnamefont {N.}~\bibnamefont {Hayashi}},
  \bibinfo {author} {\bibfnamefont {Y.}~\bibnamefont {Tanaka}},\ and\ \bibinfo
  {author} {\bibfnamefont {A.~A.}\ \bibnamefont {Golubov}},\ }\bibfield
  {title} {\bibinfo {title} {Model for vortex-core tunneling spectroscopy of
  chiral $p$-wave superconductors via odd-frequency pairing states},\ }\href
  {https://doi.org/10.1103/PhysRevLett.102.117003} {\bibfield  {journal}
  {\bibinfo  {journal} {Phys. Rev. Lett.}\ }\textbf {\bibinfo {volume} {102}},\
  \bibinfo {pages} {117003} (\bibinfo {year} {2009})}\BibitemShut {NoStop}%
\bibitem [{\citenamefont {Asano}\ and\ \citenamefont
  {Tanaka}(2013)}]{asano:prb2013}%
  \BibitemOpen
  \bibfield  {author} {\bibinfo {author} {\bibfnamefont {Y.}~\bibnamefont
  {Asano}}\ and\ \bibinfo {author} {\bibfnamefont {Y.}~\bibnamefont {Tanaka}},\
  }\bibfield  {title} {\bibinfo {title} {Majorana fermions and odd-frequency
  cooper pairs in a normal-metal nanowire proximity-coupled to a topological
  superconductor},\ }\href {https://doi.org/10.1103/PhysRevB.87.104513}
  {\bibfield  {journal} {\bibinfo  {journal} {Phys. Rev. B}\ }\textbf {\bibinfo
  {volume} {87}},\ \bibinfo {pages} {104513} (\bibinfo {year}
  {2013})}\BibitemShut {NoStop}%
\bibitem [{\citenamefont {Agterberg}\ \emph {et~al.}(2017)\citenamefont
  {Agterberg}, \citenamefont {Brydon},\ and\ \citenamefont
  {Timm}}]{agterberg:prl2017}%
  \BibitemOpen
  \bibfield  {author} {\bibinfo {author} {\bibfnamefont {D.~F.}\ \bibnamefont
  {Agterberg}}, \bibinfo {author} {\bibfnamefont {P.~M.~R.}\ \bibnamefont
  {Brydon}},\ and\ \bibinfo {author} {\bibfnamefont {C.}~\bibnamefont {Timm}},\
  }\bibfield  {title} {\bibinfo {title} {Bogoliubov fermi surfaces in
  superconductors with broken time-reversal symmetry},\ }\href
  {https://doi.org/10.1103/PhysRevLett.118.127001} {\bibfield  {journal}
  {\bibinfo  {journal} {Phys. Rev. Lett.}\ }\textbf {\bibinfo {volume} {118}},\
  \bibinfo {pages} {127001} (\bibinfo {year} {2017})}\BibitemShut {NoStop}%
\bibitem [{\citenamefont {Kim}\ \emph {et~al.}(2021)\citenamefont {Kim},
  \citenamefont {Kobayashi},\ and\ \citenamefont {Asano}}]{kim:jpsj2021}%
  \BibitemOpen
  \bibfield  {author} {\bibinfo {author} {\bibfnamefont {D.}~\bibnamefont
  {Kim}}, \bibinfo {author} {\bibfnamefont {S.}~\bibnamefont {Kobayashi}},\
  and\ \bibinfo {author} {\bibfnamefont {Y.}~\bibnamefont {Asano}},\ }\bibfield
   {title} {\bibinfo {title} {Quasiparticle on bogoliubov fermi surface and
  odd-frequency cooper pair},\ }\href {https://doi.org/10.7566/JPSJ.90.104708}
  {\bibfield  {journal} {\bibinfo  {journal} {Journal of the Physical Society
  of Japan}\ }\textbf {\bibinfo {volume} {90}},\ \bibinfo {pages} {104708}
  (\bibinfo {year} {2021})},\ \Eprint
  {https://arxiv.org/abs/https://doi.org/10.7566/JPSJ.90.104708}
  {https://doi.org/10.7566/JPSJ.90.104708} \BibitemShut {NoStop}%
\bibitem [{\citenamefont {Kuzmanovski}\ \emph {et~al.}(2020)\citenamefont
  {Kuzmanovski}, \citenamefont {Souto},\ and\ \citenamefont
  {Balatsky}}]{kuzmanovski:prb2020}%
  \BibitemOpen
  \bibfield  {author} {\bibinfo {author} {\bibfnamefont {D.}~\bibnamefont
  {Kuzmanovski}}, \bibinfo {author} {\bibfnamefont {R.~S.}\ \bibnamefont
  {Souto}},\ and\ \bibinfo {author} {\bibfnamefont {A.~V.}\ \bibnamefont
  {Balatsky}},\ }\bibfield  {title} {\bibinfo {title} {Odd-frequency
  superconductivity near a magnetic impurity in a conventional
  superconductor},\ }\href {https://doi.org/10.1103/PhysRevB.101.094505}
  {\bibfield  {journal} {\bibinfo  {journal} {Phys. Rev. B}\ }\textbf {\bibinfo
  {volume} {101}},\ \bibinfo {pages} {094505} (\bibinfo {year}
  {2020})}\BibitemShut {NoStop}%
\bibitem [{\citenamefont {Perrin}\ \emph {et~al.}(2020)\citenamefont {Perrin},
  \citenamefont {Santos}, \citenamefont {M\'enard}, \citenamefont {Brun},
  \citenamefont {Cren}, \citenamefont {Civelli},\ and\ \citenamefont
  {Simon}}]{perrin:prl2020}%
  \BibitemOpen
  \bibfield  {author} {\bibinfo {author} {\bibfnamefont {V.}~\bibnamefont
  {Perrin}}, \bibinfo {author} {\bibfnamefont {F.~L.~N.}\ \bibnamefont
  {Santos}}, \bibinfo {author} {\bibfnamefont {G.~C.}\ \bibnamefont
  {M\'enard}}, \bibinfo {author} {\bibfnamefont {C.}~\bibnamefont {Brun}},
  \bibinfo {author} {\bibfnamefont {T.}~\bibnamefont {Cren}}, \bibinfo {author}
  {\bibfnamefont {M.}~\bibnamefont {Civelli}},\ and\ \bibinfo {author}
  {\bibfnamefont {P.}~\bibnamefont {Simon}},\ }\bibfield  {title} {\bibinfo
  {title} {Unveiling odd-frequency pairing around a magnetic impurity in a
  superconductor},\ }\href {https://doi.org/10.1103/PhysRevLett.125.117003}
  {\bibfield  {journal} {\bibinfo  {journal} {Phys. Rev. Lett.}\ }\textbf
  {\bibinfo {volume} {125}},\ \bibinfo {pages} {117003} (\bibinfo {year}
  {2020})}\BibitemShut {NoStop}%
\bibitem [{\citenamefont {Tanaka}\ \emph {et~al.}(2005)\citenamefont {Tanaka},
  \citenamefont {Asano}, \citenamefont {Golubov},\ and\ \citenamefont
  {Kashiwaya}}]{tanaka:prb2005}%
  \BibitemOpen
  \bibfield  {author} {\bibinfo {author} {\bibfnamefont {Y.}~\bibnamefont
  {Tanaka}}, \bibinfo {author} {\bibfnamefont {Y.}~\bibnamefont {Asano}},
  \bibinfo {author} {\bibfnamefont {A.~A.}\ \bibnamefont {Golubov}},\ and\
  \bibinfo {author} {\bibfnamefont {S.}~\bibnamefont {Kashiwaya}},\ }\bibfield
  {title} {\bibinfo {title} {Anomalous features of the proximity effect in
  triplet superconductors},\ }\href
  {https://doi.org/10.1103/PhysRevB.72.140503} {\bibfield  {journal} {\bibinfo
  {journal} {Phys. Rev. B}\ }\textbf {\bibinfo {volume} {72}},\ \bibinfo
  {pages} {140503} (\bibinfo {year} {2005})}\BibitemShut {NoStop}%
\bibitem [{\citenamefont {Asano}\ \emph {et~al.}(2011)\citenamefont {Asano},
  \citenamefont {Golubov}, \citenamefont {Fominov},\ and\ \citenamefont
  {Tanaka}}]{asano:prl2011}%
  \BibitemOpen
  \bibfield  {author} {\bibinfo {author} {\bibfnamefont {Y.}~\bibnamefont
  {Asano}}, \bibinfo {author} {\bibfnamefont {A.~A.}\ \bibnamefont {Golubov}},
  \bibinfo {author} {\bibfnamefont {Y.~V.}\ \bibnamefont {Fominov}},\ and\
  \bibinfo {author} {\bibfnamefont {Y.}~\bibnamefont {Tanaka}},\ }\bibfield
  {title} {\bibinfo {title} {Unconventional surface impedance of a normal-metal
  film covering a spin-triplet superconductor due to odd-frequency cooper
  pairs},\ }\href {https://doi.org/10.1103/PhysRevLett.107.087001} {\bibfield
  {journal} {\bibinfo  {journal} {Phys. Rev. Lett.}\ }\textbf {\bibinfo
  {volume} {107}},\ \bibinfo {pages} {087001} (\bibinfo {year}
  {2011})}\BibitemShut {NoStop}%
\bibitem [{\citenamefont {Suzuki}\ and\ \citenamefont
  {Asano}(2014)}]{suzuki:prb2014}%
  \BibitemOpen
  \bibfield  {author} {\bibinfo {author} {\bibfnamefont {S.-I.}\ \bibnamefont
  {Suzuki}}\ and\ \bibinfo {author} {\bibfnamefont {Y.}~\bibnamefont {Asano}},\
  }\bibfield  {title} {\bibinfo {title} {Paramagnetic instability of small
  topological superconductors},\ }\href
  {https://doi.org/10.1103/PhysRevB.89.184508} {\bibfield  {journal} {\bibinfo
  {journal} {Phys. Rev. B}\ }\textbf {\bibinfo {volume} {89}},\ \bibinfo
  {pages} {184508} (\bibinfo {year} {2014})}\BibitemShut {NoStop}%
\bibitem [{\citenamefont {Asano}\ and\ \citenamefont
  {Sasaki}(2015)}]{asano:prb2015}%
  \BibitemOpen
  \bibfield  {author} {\bibinfo {author} {\bibfnamefont {Y.}~\bibnamefont
  {Asano}}\ and\ \bibinfo {author} {\bibfnamefont {A.}~\bibnamefont {Sasaki}},\
  }\bibfield  {title} {\bibinfo {title} {Odd-frequency cooper pairs in two-band
  superconductors and their magnetic response},\ }\href
  {https://doi.org/10.1103/PhysRevB.92.224508} {\bibfield  {journal} {\bibinfo
  {journal} {Phys. Rev. B}\ }\textbf {\bibinfo {volume} {92}},\ \bibinfo
  {pages} {224508} (\bibinfo {year} {2015})}\BibitemShut {NoStop}%
\bibitem [{\citenamefont {Eilenberger}(1968)}]{eilenberger:zphys1968}%
  \BibitemOpen
  \bibfield  {author} {\bibinfo {author} {\bibfnamefont {G.}~\bibnamefont
  {Eilenberger}},\ }\bibfield  {title} {\bibinfo {title} {Transformation of
  gorkov's equation for type ii superconductors into transport-like
  equations},\ }\href {https://doi.org/10.1007/BF01379803} {\bibfield
  {journal} {\bibinfo  {journal} {Zeitschrift f{\"u}r Physik A Hadrons and
  nuclei}\ }\textbf {\bibinfo {volume} {214}},\ \bibinfo {pages} {195}
  (\bibinfo {year} {1968})}\BibitemShut {NoStop}%
\bibitem [{\citenamefont {Larkin}\ and\ \citenamefont
  {Ovchinnikov}(1969)}]{larkin:jetp1969}%
  \BibitemOpen
  \bibfield  {author} {\bibinfo {author} {\bibfnamefont {A.~I.}\ \bibnamefont
  {Larkin}}\ and\ \bibinfo {author} {\bibfnamefont {Y.~N.}\ \bibnamefont
  {Ovchinnikov}},\ }\bibfield  {title} {\bibinfo {title} {Quasiclassical method
  in the theory of superconductivity},\ }\href@noop {} {\bibfield  {journal}
  {\bibinfo  {journal} {Sov. Phys. JETP}\ }\textbf {\bibinfo {volume} {28}},\
  \bibinfo {pages} {1200} (\bibinfo {year} {1969})}\BibitemShut {NoStop}%
\bibitem [{\citenamefont {Rouco}\ \emph {et~al.}(2019)\citenamefont {Rouco},
  \citenamefont {Tokatly},\ and\ \citenamefont {Bergeret}}]{rouco:prb2019}%
  \BibitemOpen
  \bibfield  {author} {\bibinfo {author} {\bibfnamefont {M.}~\bibnamefont
  {Rouco}}, \bibinfo {author} {\bibfnamefont {I.~V.}\ \bibnamefont {Tokatly}},\
  and\ \bibinfo {author} {\bibfnamefont {F.~S.}\ \bibnamefont {Bergeret}},\
  }\bibfield  {title} {\bibinfo {title} {Spectral properties and quantum phase
  transitions in superconducting junctions with a ferromagnetic link},\ }\href
  {https://doi.org/10.1103/PhysRevB.99.094514} {\bibfield  {journal} {\bibinfo
  {journal} {Phys. Rev. B}\ }\textbf {\bibinfo {volume} {99}},\ \bibinfo
  {pages} {094514} (\bibinfo {year} {2019})}\BibitemShut {NoStop}%
\bibitem [{\citenamefont {Asano}\ \emph {et~al.}(2014)\citenamefont {Asano},
  \citenamefont {Fominov},\ and\ \citenamefont {Tanaka}}]{asano:prb2014}%
  \BibitemOpen
  \bibfield  {author} {\bibinfo {author} {\bibfnamefont {Y.}~\bibnamefont
  {Asano}}, \bibinfo {author} {\bibfnamefont {Y.~V.}\ \bibnamefont {Fominov}},\
  and\ \bibinfo {author} {\bibfnamefont {Y.}~\bibnamefont {Tanaka}},\
  }\bibfield  {title} {\bibinfo {title} {Consequences of bulk odd-frequency
  superconducting states for the classification of cooper pairs},\ }\href
  {https://doi.org/10.1103/PhysRevB.90.094512} {\bibfield  {journal} {\bibinfo
  {journal} {Phys. Rev. B}\ }\textbf {\bibinfo {volume} {90}},\ \bibinfo
  {pages} {094512} (\bibinfo {year} {2014})}\BibitemShut {NoStop}%
\bibitem [{\citenamefont {Higashitani}(2014)}]{higashitani:prb2014}%
  \BibitemOpen
  \bibfield  {author} {\bibinfo {author} {\bibfnamefont {S.}~\bibnamefont
  {Higashitani}},\ }\bibfield  {title} {\bibinfo {title} {Odd-frequency pairing
  effect on the superfluid density and the pauli spin susceptibility in
  spatially nonuniform spin-singlet superconductors},\ }\href
  {https://doi.org/10.1103/PhysRevB.89.184505} {\bibfield  {journal} {\bibinfo
  {journal} {Phys. Rev. B}\ }\textbf {\bibinfo {volume} {89}},\ \bibinfo
  {pages} {184505} (\bibinfo {year} {2014})}\BibitemShut {NoStop}%
\bibitem [{\citenamefont {Abrikosov}\ \emph {et~al.}(1975)\citenamefont
  {Abrikosov}, \citenamefont {Gor{\textquoteright}kov},\ and\ \citenamefont
  {Dzyaloshinski}}]{agd}%
  \BibitemOpen
  \bibfield  {author} {\bibinfo {author} {\bibfnamefont {A.~A.}\ \bibnamefont
  {Abrikosov}}, \bibinfo {author} {\bibfnamefont {L.~P.}\ \bibnamefont
  {Gor{\textquoteright}kov}},\ and\ \bibinfo {author} {\bibfnamefont {I.~E.}\
  \bibnamefont {Dzyaloshinski}},\ }\href@noop {} {\emph {\bibinfo {title}
  {{Methods of Quantum Field Theory in Statistical Physics}}}}\ (\bibinfo
  {publisher} {Dover Publications},\ \bibinfo {address} {New York},\ \bibinfo
  {year} {1975})\BibitemShut {NoStop}%
\bibitem [{\citenamefont {Schopohl}(1998)}]{schopohl:arxiv1998}%
  \BibitemOpen
  \bibfield  {author} {\bibinfo {author} {\bibfnamefont {N.}~\bibnamefont
  {Schopohl}},\ }\bibfield  {title} {\bibinfo {title} {Transformation of the
  eilenberger equations of superconductivity to a scalar riccati equation},\
  }\href@noop {} {\bibfield  {journal} {\bibinfo  {journal} {arXiv:cond-mat}\
  ,\ \bibinfo {pages} {9804064}} (\bibinfo {year} {1998})}\BibitemShut
  {NoStop}%
\bibitem [{\citenamefont {Higashitani}(1997)}]{higashitani:jpsj1997}%
  \BibitemOpen
  \bibfield  {author} {\bibinfo {author} {\bibfnamefont {S.}~\bibnamefont
  {Higashitani}},\ }\bibfield  {title} {\bibinfo {title} {Mechanism of
  paramagnetic meissner effect in high-temperature superconductors},\ }\href
  {https://doi.org/10.1143/JPSJ.66.2556} {\bibfield  {journal} {\bibinfo
  {journal} {Journal of the Physical Society of Japan}\ }\textbf {\bibinfo
  {volume} {66}},\ \bibinfo {pages} {2556} (\bibinfo {year} {1997})},\ \Eprint
  {https://arxiv.org/abs/https://doi.org/10.1143/JPSJ.66.2556}
  {https://doi.org/10.1143/JPSJ.66.2556} \BibitemShut {NoStop}%
\bibitem [{\citenamefont {Walter}\ \emph {et~al.}(1998)\citenamefont {Walter},
  \citenamefont {Prusseit}, \citenamefont {Semerad}, \citenamefont {Kinder},
  \citenamefont {Assmann}, \citenamefont {Huber}, \citenamefont {Burkhardt},
  \citenamefont {Rainer},\ and\ \citenamefont {Sauls}}]{walter:prl1998}%
  \BibitemOpen
  \bibfield  {author} {\bibinfo {author} {\bibfnamefont {H.}~\bibnamefont
  {Walter}}, \bibinfo {author} {\bibfnamefont {W.}~\bibnamefont {Prusseit}},
  \bibinfo {author} {\bibfnamefont {R.}~\bibnamefont {Semerad}}, \bibinfo
  {author} {\bibfnamefont {H.}~\bibnamefont {Kinder}}, \bibinfo {author}
  {\bibfnamefont {W.}~\bibnamefont {Assmann}}, \bibinfo {author} {\bibfnamefont
  {H.}~\bibnamefont {Huber}}, \bibinfo {author} {\bibfnamefont
  {H.}~\bibnamefont {Burkhardt}}, \bibinfo {author} {\bibfnamefont
  {D.}~\bibnamefont {Rainer}},\ and\ \bibinfo {author} {\bibfnamefont {J.~A.}\
  \bibnamefont {Sauls}},\ }\bibfield  {title} {\bibinfo {title}
  {Low-temperature anomaly in the penetration depth of
  ${\mathrm{yba}}_{2}{\mathrm{cu}}_{3}{O}_{7}$ films: Evidence for andreev
  bound states at surfaces},\ }\href
  {https://doi.org/10.1103/PhysRevLett.80.3598} {\bibfield  {journal} {\bibinfo
   {journal} {Phys. Rev. Lett.}\ }\textbf {\bibinfo {volume} {80}},\ \bibinfo
  {pages} {3598} (\bibinfo {year} {1998})}\BibitemShut {NoStop}%
\bibitem [{\citenamefont {Suzuki}\ and\ \citenamefont
  {Asano}(2015)}]{suzuki:prb2015}%
  \BibitemOpen
  \bibfield  {author} {\bibinfo {author} {\bibfnamefont {S.-I.}\ \bibnamefont
  {Suzuki}}\ and\ \bibinfo {author} {\bibfnamefont {Y.}~\bibnamefont {Asano}},\
  }\bibfield  {title} {\bibinfo {title} {Effects of surface roughness on the
  paramagnetic response of small unconventional superconductors},\ }\href
  {https://doi.org/10.1103/PhysRevB.91.214510} {\bibfield  {journal} {\bibinfo
  {journal} {Phys. Rev. B}\ }\textbf {\bibinfo {volume} {91}},\ \bibinfo
  {pages} {214510} (\bibinfo {year} {2015})}\BibitemShut {NoStop}%
\bibitem [{\citenamefont {Eilenberger}(1966)}]{eilenberger:zphys1966}%
  \BibitemOpen
  \bibfield  {author} {\bibinfo {author} {\bibfnamefont {G.}~\bibnamefont
  {Eilenberger}},\ }\bibfield  {title} {\bibinfo {title} {General approximation
  method for the free energy functional of superconducting alloys},\ }\href
  {https://doi.org/10.1007/BF01327140} {\bibfield  {journal} {\bibinfo
  {journal} {Zeitschrift f{\"u}r Physik}\ }\textbf {\bibinfo {volume} {190}},\
  \bibinfo {pages} {142} (\bibinfo {year} {1966})}\BibitemShut {NoStop}%
\end{thebibliography}

\end{document}